\documentclass[useAMS,usenatbib]{mn2e}
\usepackage{epsfig}
\usepackage{afterpage}
\usepackage{subfigure}
\usepackage{longtable}
\usepackage{rotating}
\usepackage{lscape}
\usepackage{times}
\setlongtables
\def\gtrsim{\mathrel{\hbox{\rlap{\hbox{\lower4pt\hbox{$\sim$}}}\hbox{$>$}}}}

\title[Environmental dependence of AGN activity in the supercluster A901/2]{Environmental dependence of AGN activity in the supercluster A901/2}
\author[R. Gilmour et al.]{R. Gilmour$^{1,2}$\thanks{e-mail rgilmour@eso.org}
, M. E. Gray$^{3}$, O. Almaini$^{3}$, P. Best$^{1}$, C. Wolf$^4$, K. Meisenheimer$^5$, \newauthor 
C.Papovich$^6$ and E. Bell$^5$
 \\
$^{1}$Scottish Universities Physics Alliance, Institute for Astronomy, Royal Observatory, Blackford Hill, Edinburgh, EH9 3HJ, UK\\
$^{2}$European Southern Observatory, Alonso de Cordova 3107, Vitacura, Casilla 19001, Santiago 19, Chile\\
$^{3}$School of Physics and Astronomy, University of Nottingham, University Park, Nottingham,  NG7 2RD, UK\\
$^4$ Department of Physics, Denys Wilkinson Bldg., University of Oxford, Keble Road, Oxford, OX1 3RH, UK\\
$^5$ Max--Plank--Institut f\"ur Astronomie, K\"onigstuhl 17, D-69117 Heidelberg, Germany\\
$^6$ Steward Observatory, The University of Arizona, 933 North Cherry Avenue, Tuscon, AZ 85721, USA
}
\begin{document}

\date{Accepted . Received ; in original form }

\pagerange{\pageref{firstpage}--\pageref{lastpage}} \pubyear{2007}

\maketitle

\label{firstpage}

\begin{abstract}

We present XMM data for the supercluster A901/2, at $z\sim0.17$, which
is combined with deep imaging and 17-band photometric redshifts (from
the COMBO-17 survey), 2dF spectra and Spitzer 24$\mu$m data, to
identify AGN in the supercluster. The 90ksec XMM image contains 139
point sources, of which 11 are identified as supercluster AGN with
$L_{X(0.5-7.5keV)} > 1.7\times 10^{41}$erg cm$^{-2}$s$^{-1}$. The host
galaxies have $M_{\rm R} < -20$ and only 2 of 8 sources with spectra
could have been identified as AGN by the detected optical emission
lines. Using a large sample of 795 supercluster galaxies we define
control samples of massive galaxies with no detected AGN. The local
environments of the AGN and control samples differ at $\gtrsim 98$ per
cent significance. The AGN host galaxies lie predominantly in areas of
moderate projected galaxy density and with more local blue galaxies
than the control sample, with the exception of one very bright Type I
AGN very near the centre of a cluster. These environments are similar
to, but not limited to, cluster outskirts and blue groups. Despite the
large number of potential host galaxies, no AGN are found in regions
with the highest galaxy density (excluding some cluster cores where
emission from the ICM obscures moderate luminosity AGN).  AGN are also
absent from the areas with lowest galaxy denstiy. We conclude that the
prevalence of cluster AGN is linked to their environment.

\end{abstract}


\section{Introduction}

The properties and evolution of galaxies are known to be strongly
linked to their external environment. In particular, populations of
galaxies in clusters are strikingly different to those in the field, as
shown by the morphology--density relation (\citealt{Oemler},
\citealt{DresslerMorph}) and dramatic changes in star-formation rates
(e.g. \citealt{2dfCluster} and \citealt{SLOANcluster}).  It appears
that galaxies change significantly as they join denser environments
such as groups and clusters.

It is increasingly evident that many of the changes in galaxy
properties between cluster cores and the field are triggered in
intermediate density environments, and that a distinction between
field and cluster populations is overly simplistic. For example,
\cite{Wake} find that galaxy colour is a function of local rather
than extended galaxy density, and \cite{2dfCluster} and
\cite{SLOANcluster} find the same result for star-formation
rate. \cite{Wolf05} find that dusty star-forming galaxies are
generally found in moderate density environments.

Various processes have been suggested to account for the rapid
transformation of cluster galaxies, from local effects such as mergers
(e.g. \citealt{Mihos2}) and repeated close gravitational encounters
(e.g. \citealt{Moore}) to the large scale effects like the tidal field
(e.g. \citealt{Byrd}) and the intra-cluster medium, via ram-pressure stripping
(e.g. \cite{Abadi}) or `strangulation' (e.g. \citealt{Larson}).

The processes which affect galaxy properties may also, directly or
indirectly, affect the accretion onto the central black hole found in
most, if not all, galaxies with a stellar bulge (e.g
\citealt{Magorrian}). Both local and large-scale processes which may
affect cluster galaxies also have the potential to affect the
distribution of gas in the galaxies, and hence may trigger or suppress
AGN activity. Recently large and moderate sized surveys have begun to
shed light on the local and extended environments of AGN, and produce
observational evidence for some of these processes.

The first evidence of a suppression of AGN in the cores of galaxy
clusters was found in the optical survey of \cite{DresslerSurvey}, although
a large source of bias was suggested by \cite{Edge}. Further optical
surveys (such as \citealt{Coldwell1} and
\citealt{Kauffmann2}) have also found a deficit of luminous
AGN in dense regions. However \cite{Miller} find that the fraction of
luminous galaxies with AGN is independent of galaxy density, a
conclusion also drawn from the auto-correlations of AGN and galaxies
presented by \cite{Wake}. Surprisingly, some of these studies use the
same datasets, but draw contrasting conclusions, probably due to
different AGN selection techniques.

The picture is further complicated when detections in other wavebands
are considered. In contrast to the optical results, many radio studies
(e.g. \citealt{BestRadio}, \citealt{MillerRadio}, \citealt{Barr},
\citealt{Reddy}, and \citealt{Best}) show an increase in radio-loud
AGN activity in galaxy clusters, at a range of redshifts and in both
relaxed and merging systems. \cite{Best} find that the majority of
radio-loud AGN in the densest regions are not emission-line sources,
and so may be missed by optical studies.

However, radio-loud AGN are not representative of AGN as a whole, and
optical studies are prone to selection effects: studies of X-ray
emission are an alternative method to remove some selection effects,
and to detect a larger population of AGN. Indeed, \cite{Martini2} find
that only four of at least 35 X-ray detected AGN in a sample of
clusters have optical spectral signatures of AGN activity. The
majority of studies in the X-ray have focused on galaxy clusters,
which have the advantage of a large number of galaxies and high
density, but are complicated by the X-ray emission from the
intra-cluster medium. This tends to mask any detections of AGN in the
very centre of the cluster, in particular in the central galaxy, if
one exists, which is often a massive elliptical and radio loud
(e.g. \citealt{Peres}, \citealt{Bestastroph}). The exclusion of such
galaxies may in fact be an advantage as they are in very different
environments from the other cluster galaxies, frequently lying in the
centre of the potential well. This unusual environment probably has a
very different effect on AGN activity compared to the other cluster
members, and it is therefore preferable to distinguish between
central and normal galaxies when evaluating the environments of AGN.

Statistical studies of point sources in the fields of galaxy clusters
have found numerous clusters that have more luminous point sources than
expected from a non-cluster field (e.g. \citealt{Henry},
\citealt{Lazzati}, \citealt{Molnar}, \citealt{Cappi},
\citealt{Olivia}, \citealt{Cappelluti} and \citealt{Ruderman} and
Gilmour et al. in prep.) and which are therefore likely to contain
AGN.  Others studies find clusters which appear to have no excess of
sources (e.g. \cite{Molnar}, \citeauthor{Champ1} \nocite{Champ2}2004a
and 2004b), and therefore no detectable (generally moderate
luminosity) AGN.

\cite{Martini2} confirm spectroscopically that eight low-redshift
clusters each contain between 2 and 10 X-ray sources with $L_X
>10^{41}$erg s$^{-1}$, the majority of which are AGN without optical
emission lines. This corresponds to 5 per cent of galaxies with
$M_{\rm R} < -20$ hosting AGN with $L_X > 10^{41}$erg s$^{-1}$, which
is a far higher AGN fraction than previously determined from optical
surveys. Recent results \citep{Martini_pos} show that the radial
distribution of the fainter AGN in this sample follows that of the
cluster population, but the more luminous AGN are found preferentially
in the central regions. This result is in agreement
with a recent statistical survey of 18 clusters \citep{Branchesi}.

Given that many galaxy transformations occur in intermediate density
environments, it may be that AGN are also altered by the
host galaxy environment.  To understand the links between AGN activity and
their extended environment it is desirable to look beyond galaxy
clusters as a single entity, and instead determine the effect of local
($\sim 100$ kpc) and large scale ($\sim 1$ Mpc) environment, from the
field through groups and cluster outskirts to the cluster cores.

Superclusters are ideal testbeds for such a study as they consist of a
large number of galaxies in a range of environments, but at the same
epoch. The correlations between environment and AGN properties can
therefore be studied in one field, without complications due to galaxy
or AGN evolution. For example, the AGN population in galaxy groups 
can be compared to that in cluster outskirts of similar local galaxy
density to distinguish between local and large scale environments. In
addition superclusters contain both disturbed and relaxed
regions, which may affect AGN in different ways.

This paper presents the results of investigations into AGN in the
supercluster A901/2, which has extensive multi-wavelength imaging and
spectroscopy, summarised in Section \ref{A901-data}. The data
reduction and identification of the supercluster AGN are described in
Sections \ref{A901-xray} and \ref{A901-match}.  The properties of the
AGN and their host galaxies are investigated, along with the
environments in which they are found, and the environments of the AGN
hosts are compared to other supercluster galaxies in Section
\ref{A901-analysis}. The results are summarised in Section
\ref{A901-conclusions}. Details of individual AGN candidates are given
in the Appendix. Throughout the paper the cosmological parameters
$\Omega_m, \Omega_\Lambda$ and $H_0$ are set to 0.3, 0.7 and 70 km
s$^{-1}$ Mpc$^{-1}$, and all COMBO-17 absolute magnitudes are
converted to this cosmology.

\section{The supercluster A901/2}\label{A901-data}
\subsection{Optical data}

The supercluster consisting of Abell 901 and Abell 902 (A901/2), first
identified by \cite{Abell}, is ideal for a study of the effect of
environment on AGN due to the low redshift ($\sim 0.17$) and wealth of
optical data available. It is one of the fields in the
COMBO-17 survey (Classifying Objects by Medium-Band Observations in 17
Filters, \citealt{Wolf03}), and in addition 2dF spectra are
available for 282 supercluster galaxies, from observations with the
two degree field (2dF) spectrograph on the Anglo--Australian telescope.

The COMBO-17 survey used the Wide Field Imager (WFI) at the MPG/ESO
2.2 m telescope to obtain images of a $0.56 \times
0.55$ degree field with a pixel size of $0.238\arcsec$.  Images were
taken in 5 broad and 12 narrow band filters and matched to a set of
template spectra to determine photometric redshifts ($z_{\rm phot}$).
Reliable photometric redshifts were found for the $\sim$18000 objects
with $m_{\rm R} < 24 $, with errors of order $\sigma_z/(1+z)<0.01$
(which is comparable to the velocity dispersion of the supercluster)
at $m_{\rm R} < 20 $, and $\sigma_z/(1+z)<0.02$ for $m_{\rm R} < 23 $
(\cite{Wolf05}, hereafter WGM05). The accuracy of the photometric
redshifts when compared to the available spectroscopic redshifts is
shown by \citeauthor{Wolf} to be good, such that it is possible to
select a magnitude-limited supercluster sample with minimal
contamination from interlopers and only a few percent loss of true
supercluster galaxies.

A cut of $0.155 < z_{\rm phot} < 0.185$ gives 
795 galaxies with total absolute V band magnitude $<-17$, which are
used in WGM05 and \cite{Lane}.  This large
sample makes it possible to determine very accurately the distribution
and properties of the galaxies in A901/2 \citep{Gray}.

\begin{figure}
\includegraphics[angle=270,width=\columnwidth]{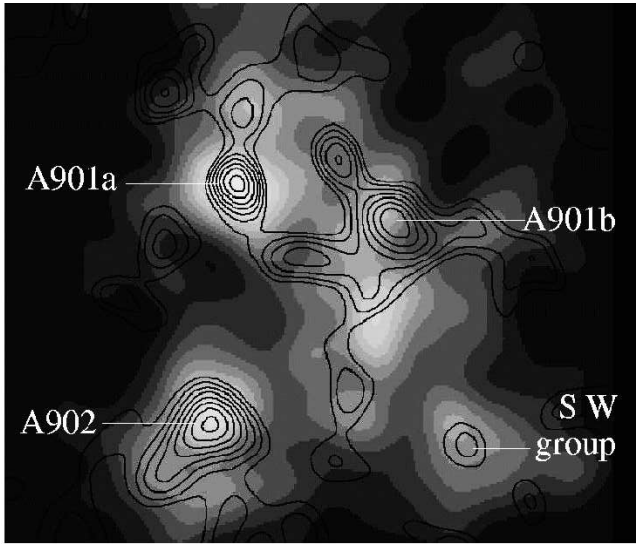}
\caption[Galaxy number density and Dark Matter in A901/2 ]{ Galaxy
number density in A901 (grey-scale) and supercluster dark matter
density from 2D weak lensing analysis by \cite{Gray02}
(contours). Galaxies are selected in the range $0.155 < z_{\rm phot} <
0.185$. The optically identified clusters and group are marked.  The
image is $\sim 30 \arcmin$ square and the top left corner is
North-East.
\label{A901dm}}
\end{figure}

The optically identified structure of A901/2 is shown in Figure
\ref{A901dm}. A901 consists of two dark matter halos of comparable
size, A901a and A901b, each with a massive galaxy in the centre. A901a
contains far more galaxies and is far more concentrated than A901b. A
tail of smaller, bluer galaxies extends south of A901b towards A902,
which is a more optically diffuse cluster. There is also a group of
galaxies in the south-west corner, and optical data and 3D weak
lensing analysis \citep{Taylor} has identified a cluster at redshift
$\sim 0.5$ almost directly behind A902.  It is clear that the
supercluster contains a wide range of environments with differing
ratios of dark to optically visible matter. The effect of these
environments on the galaxy star-formation rate has been investigated
by \cite{Gray}, who found that the proportion of galaxies that are
star-forming is a strong function of local dark-matter density, with
far less star-formation in galaxies in denser regions.  In addition,
the WGM05 study found a population of dusty star-forming galaxies which
preferentially exist in medium galaxy-density environments, avoiding both the
field and the cluster cores.

2dF spectra are available for 282 of the brightest galaxies in the
supercluster, and cover the wavelength range 3900--6000A (Gray et
al. in prep.). The COMBO-17 SEDs and 2dF spectra are not sufficient to
compile a sample of AGN as many AGN are optically obscured. In
comparison, X-ray samples are far more complete (see for example
\citealt{Martini2} and \citealt{Szokoly}) but suffer from confusion
with heavily star-forming galaxies. Combining an X-ray source list
with the 2dF spectra and COMBO-17 data can help identify supercluster
X-ray sources and distinguish between X-ray emission from AGN and that
from other sources such as star-formation and populations of low mass X-ray
binaries. Comparing the positions of AGN hosts with the other
identified supercluster galaxies will determine whether AGN activity
is enhanced or suppressed in a range of environments.

\subsection{Infra-red data}
This field is currently being surveyed using MIPS (Multi-band Imaging
Photometer for Spitzer) on Spitzer. An early release catalogue of the 24
micron sources in this field (1/7 of the final data, from Bell and
Papovich, private communication) will help in the identification of some
AGN, as shown in Section \ref{SpitzerSection}.

\subsection{X-ray data}

The A901/2 supercluster region had previously been observed for $\sim
0.4$ ksec as part of the ROSAT All Sky Survey \citep{EbelingRASS} and
in addition with the ROSAT High Resolution Imager for $\sim 12$ksec.
\cite{Schindler} found seven sources in the field, two of which
coincide with A901a and A901b. The bright emission coincident with
A901a was found to be a point source.

In this paper a new, deep (90ksec) XMM-Newton image of the
supercluster, obtained in 2003, is presented. By combining the deep
X-ray image and the optical data a sample of AGN in the supercluster
are selected and analysed.

\section{X-ray data reduction}\label{A901-xray}

\subsection{Data reduction}\label{X-raydata}

A 90ksec XMM image of A901/2 was taken on 6th/7th May 2003 using the
three EPIC cameras (MOS1, MOS2, PN) and a thin filter.  The level 1
data were taken from the supplied pipeline products, and reduced with
SAS v5.4 and the calibration files.  The data were filtered for bad
pixels, the standard good patterns of 0--12 and XMMEA\_EM or XMMEA\_EP
and energy between 0.5 and 7.5 keV.  Removing times when the count
rate was $> 0.2$ counts s$^{-1}$ for MOS1 and MOS2 detectors and $>
0.67$ counts s$^{-1}$ for PN resulted in an exposure time of $\sim
67$ksec for MOS and $\sim 61$ksec for PN, and removed all episodes of
significant flaring.

\subsection{Source detection}\label{SourceDetection}

Sources were detected using {\sc WAVDETECT} \citep{Wavdetect} on 600 x
600 pixel unvignetted full-band images for each detector. The images
and the corresponding exposure maps had a pixel size of 4.1\arcsec. A
mask was created for each detector, which removed areas with less than
25 per cent of the maximum exposure or an exposure map gradient of over 0.4
for MOS or 0.03 for PN. Three areas of streaking were removed by hand
in the PN mask.

102 sources were detected in the MOS1 image, 96 in MOS2 and 128 in PN. The
total number of unique sources detected, without applying any cut on source
significance, was 150 (of which 64 were detected in all three images, 33 in
two and 53 in one). The vast majority of those missed in one or two images
were outside the field of view of those detectors, or only detected in the
more sensitive PN image.

A point source catalogue was constructed for each detector by
removing all detections of extended supercluster emission. As the size
and shape of the PSF is not well defined in XMM two methods were used
to determine which were point sources:
\begin{itemize}
\item{The FWHM was found for each object. As the sources become increasingly
elliptical towards the edge of the image, it was required that the
semi-minor axis had a FWHM of $< 3 $ pixels. This includes all bright
on-axis point sources, which have a FWHM of $2.2$ pixels, and allows
some margin of error for the fainter sources. This method was only
useful for moderate to bright sources.}
\item{The catalogue was compared to the results for this field from
the XCS survey (private communication, see M. Davidson, in
prep.). This survey uses a sophisticated wavelet reconstruction method
to find extended emission in XMM images. Due to problems with the raw
dataset this method could only use the MOS data in the NE quarter of
the image, whereas all data are used in the rest of the
image. Detections of extended emission are therefore less accurate in
the NE quarter. }
\end{itemize}

\begin{figure*}
\includegraphics[clip,width=\textwidth]{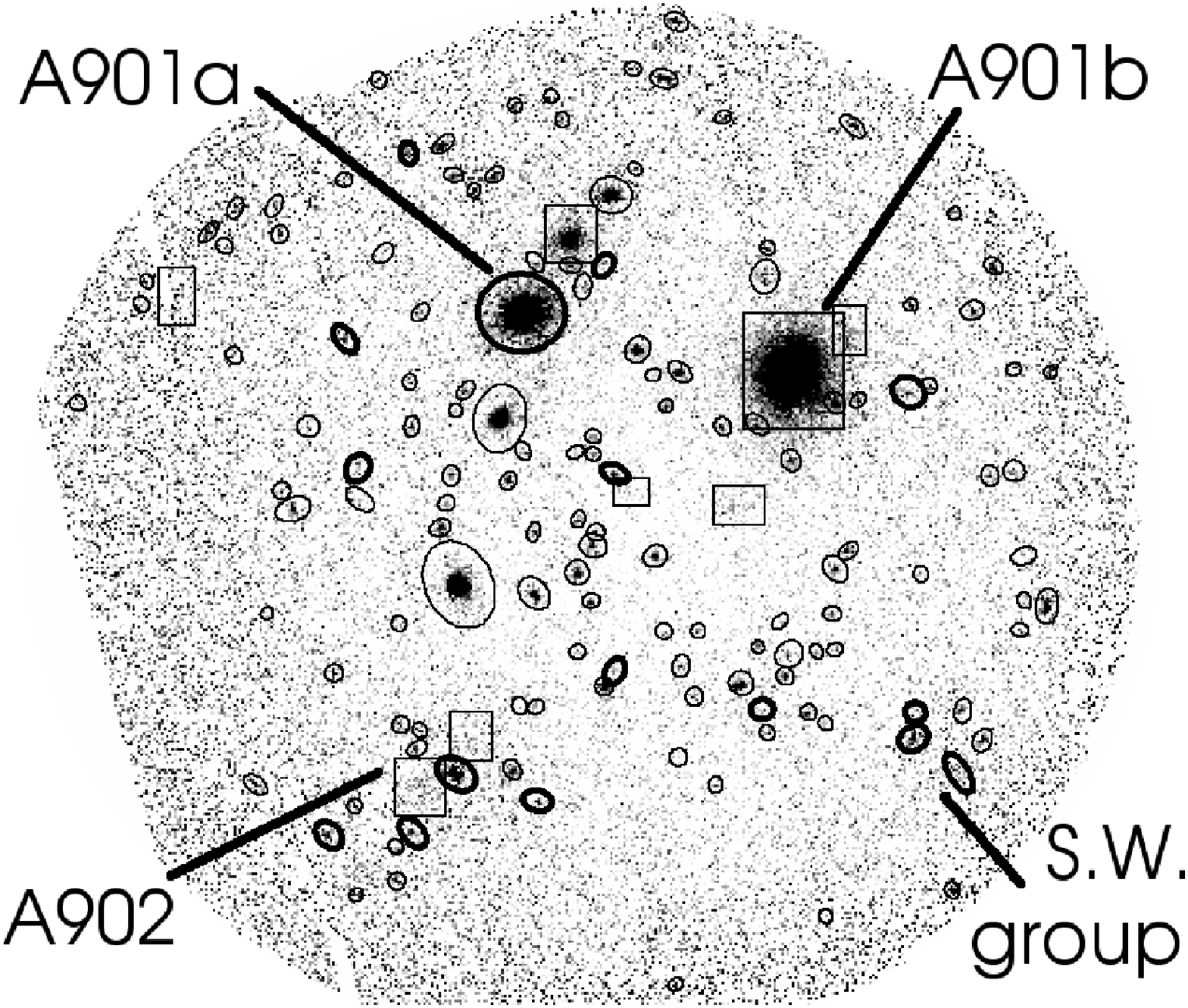}
\caption[]{The identified sources in the supercluster field. Sources
marked with a rectangle are possible extended emission. Point sources
with significance $\frac{C}{\sigma_B}> 3$ are marked with an ellipse,
and those drawn in bold are possible supercluster members. The
emission in the region of A901a is a very bright point source rather than
extended emission. The sources are plotted over a combined, vignetting
weighted image, and some sources are not visible by eye due to the
image combination and scale. The image is $\sim 30 \arcmin$ in
diameter and the top left corner is North-East.\label{xray_sources}}
\end{figure*}

The results of these methods are broadly in agreement, within the errors
described, and identified eight areas of possible supercluster emission,
shown in Figure \ref{xray_sources}. Of these, the smallest three are
likely to be artifacts as they all occur near the chip boundary of one
image only. All of these sources were removed from the catalogue. Analysis of the extended emission will be covered in Gray et al., in prep.

A further consideration is the brightest source in the field, which
has the FWHM of a point source. As this source is so bright (similar
in flux to the X-ray emission from A901b) and lies very close to the
centre of A901a, it could be concentrated cluster emission or a
cooling flow. These scenarios were ruled out by analysis of the
spectrum, which is a power law rather than thermal and the fact that
the X-ray emission is centred on a galaxy which is not the brightest
cluster galaxy and which has radio emission (from the NVSS,
\citealt{NVSS}). It is therefore concluded that this object is an AGN.

\subsection{Point source properties}\label{point_src_prop}

The reality, position and positional error of the sources were found
by comparing the sources found in each of the three detectors. For
sources that were detected in more than one image the position was
defined as the midpoint of the two closest positions for that
source. (In the most common case of sources detected in all three
images this removed errors due to one detection being near a chip
boundary.)  For singly detected sources the given position was used.

The {\sc wavdetect} 1-sigma positional errors were found to be
generally less than one arcsecond even for the faintest sources. In
comparison, the separations between detections of the same source in
different images were on average 3 arcseconds, which is just less than
one pixel. This is a random error, rather than astrometric, due to the
difficulty of finding the centre of faint objects with a large pixel
size compared to the PSF, and is a better measurement of the `true'
error on the stated source position. The {\sc wavdetect} errors are
dependent on the source size and counts, and are correlated with the
distance between detections in different images. This distance is on
average 7 times the {\sc wavdetect} error for low-significance sources
(only changing to 6 times higher for high-significance
sources). Therefore, although the {\sc wavdetect} errors are
unphysically low they can be used to estimate the true error,
especially for singly detected sources.  For this reason the error on
the source position was given by the larger of the following three
measurements: the distance between the two closest detections (if they
existed), 7 times the stated {\sc wavdetect} error, or 0.5 pixel (an
error of less than half a pixel was defined as unphysical).

As a final stage, all sources that were not detected at significance
$> 3$ in at least one image were removed, where the significance is
given by $\frac{C}{\sigma_B}$, for source counts $C$ and background
counts $B$. (This means that all sources are not random fluctuations
to at least $3\sigma$ significance, but in most cases is overly
conservative as it doesn't take into account the spatial distribution
of photons within the source area). This cut removed 11 sources, and
also agrees well with the reality determined by eye. At least one of
these sources is real (at $\sim$09:56:24 -10:01:52, probably matching
a $z_{\rm phot}$=2.2 quasar in the COMBO-17 catalogue), as it was
marginally detected in two images; however it has a very large
positional error, and for consistency and accuracy was not included in
the catalogue.

The final list of 139 significant point sources, including positional
errors, is given in Table \ref{A901table} and shown in Figure \ref{xray_sources}.

\section{Finding the Supercluster Active Galactic Nuclei}\label{A901-match}
\subsection{Matching X-ray and Optical catalogues}

The COMBO-17 catalogue consists of 63776 objects detected using
SExtractor on the R-band image \citep{Wolf03}. These were matched with
the XMM point sources to identify the X-ray sources in the
supercluster. Some saturated stars and fainter objects near
diffraction spikes are not included in the COMBO-17 catalogue, so the
areas around each X-ray source were examined manually for missing
objects, and four such optical objects which could possibly match an
X-ray source were added to the catalogue.

\begin{figure}
\begin{center}
\includegraphics[clip,width=\columnwidth,angle=0]{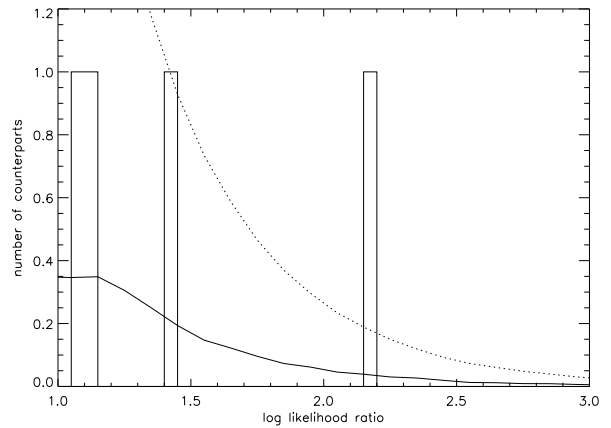}
\caption[The expected and actual distribution of likelihood ratios of
possible optical counterparts to an X-ray source]{The expected
distribution of likelihood ratios for one X-ray source based on 14000
trials (solid line), cumulative expected distribution (dashed line)
and the histogram of actual likelihood ratios for the optical objects
near this source (the vast majority of optical objects have
log(likelihood) $< 1$). One possible match is identified, with a
reliability of $\sim 0.8$.
\label{likelihood}}
\end{center}
\end{figure}

A maximum-likelihood technique was used to match the X-ray sources to
the COMBO-17 optical catalogue. Matching was performed by comparing
the value $LR_{i,j}$ (a measure of the association between two
sources, i and j, see Equation \ref{eqn1}) with the distribution of
this value for X-ray sources placed randomly within the field. Because
$LR_{i,j}$ depends on the error on the source co-ordinates, which
varies significantly in the X-ray sample, the expected distribution of
$LR_{i,j}$ was calculated for each X-ray source, $j$, by randomly
placing 14000 X-ray sources with error $\sigma_j$ over the optical
catalogue. The resulting (normalised) distribution $N(LR)_j$ gives the
probability of obtaining each likelihood ratio by chance (if source
$j$ had no optical counterpart), as shown for one source in Figure
\ref{likelihood}.

The likelihood ratio was defined (following \citealt{Mann97} and \citealt{Emma}, who use a method described in detail in \citealt{SutherlandSaunders}) as

\begin{equation}
LR_{i,j} = \frac{e^{-r_{i,j}^2/{2\sigma_j^2}}}{\sigma_j^2 N(<m_i)}\label{eqn1}
\end{equation}

\noindent where $\sigma_j$ is the positional error on X-ray source
$j$, $r_{i,j}$ the distance to optical object $i$ from X-ray source
$j$, and $N(<m_i)$ the number of optical objects brighter than object
$i$ in the r band image. This takes account of angular separation and
optical magnitude, but made no distinction between object
classification or photometric redshift.  The errors on the positions
of the optical objects were small enough to neglect compared to those
in the X-ray, and the astrometric errors were also found to be
negligible as the minimum error on the X-ray position of 0.5 pixels is
significantly larger than the astrometric error on this image.

This method treats optical quasars and galaxies of the same flux in
the same way, and does not account for the fact that quasars are rarer
and more likely to be X-ray sources. In addition, the method does not
distinguish between the brightest galaxies, which are rare and quite
likely to be X-ray sources, and stars of a similar magnitude.  These
issues are important when a faint QSO is the only match and is
assigned too low a probability. This will lead to incompleteness in
the X-ray matching, but will not affect the supercluster sample. In
addition, the probabilities will not be accurate if more than one
possible match is identified, and one of the matches is a QSO or
bright star. This can affect the supercluster sample, and in these cases
the optical classifications, locations and errors of all the possible
matches were examined in detail to determine the true source of the
X-rays.

In addition this technique compares likelihoods to the average over
the field and ignores any clustering, which gives slightly higher
likelihood values for X-ray sources in the line of sight to cluster
centres, and slightly lower for those in areas with few supercluster
galaxies. This is very unlikely to change any results, especially as
the catalogue is dominated at all optical magnitudes by
non-supercluster objects ($>80$ per cent of optical objects are not in the
supercluster at $m_{\rm R}<20$, and $ >93$ per cent at $ m_{\rm R}<24$).

The standard method of calculating the reliability of a match requires
a large sample of sources, so instead the reliability for each X-ray
-- optical pair $i,j$ was defined as the probability of not obtaining
$LR_{i,j}$ randomly,

\begin{equation}
R_{i,j} =1-\frac{\Sigma{N(LR_j > LR_{i,j})}}{14000}. 
\end{equation}

Most X-ray sources have more than one potential optical counterpart,
as well as a significant probability of having no match, such that the
reliabilities sum to $>1$. The probability that optical object $i$ is
the true counterpart to $j$, ($P_{i,j}$) and the probability that
there is no counterpart ($P_{none,j}$), given a set of possible
counterparts, $k$, are calculated following \cite{Rutledge},

\begin{equation}\label{prob1}
P_{i,j}=\frac{R_{i,j} \prod_{k\ne i}^M (1-R_{k,j})}{S} 
\end{equation}
\begin{equation}\label{prob2}
P_{none,j}=\frac{\prod_{k=1}^M(1-R_{k,j})}{S}
\end{equation}

\noindent where S is a normalisation factor so that the probabilities sum
to 1, and $M$ is the number of possible optical counterparts to the X-ray
source.

A liberal catalogue of possible and secure matches was
constructed. The thresholds used were deliberately loose as many
factors could increase the likelihood of a match on a case by case
basis. These include matching with other wavelengths, visual
inspection of the shape and centring of the X-ray point source, and
characteristics of the optical counterpart, for example if it is a
faint quasar. In addition, some of the X-ray sources have such large
errors that the chance of a random association will always be high,
even if the match appears to be excellent. Therefore all matches which
fulfill the criteria described below are included in Table
\ref{A901table}, and the additional factors which may affect the
probability of a match are listed in the `Notes' column. The possible
supercluster AGN are evaluated on a case by case basis.

For sources with only one likely optical counterpart, a match is
defined as $P_{i,j} > 0.8$, which is confirmed by visual inspection.
For sources with more than one possible counterpart the conditions for
object $i$ to be a unique counterpart given a set of options, $k$,
were $\sum_k P_{k,j} > 0.8$ and $P_{i,j} / \sum_{k \ne i}P_{k,j} >
4$. This resulted in 66 secure identifications out of 139 sources, of
which fewer than 6 are expected to be random associations.  For sources
with multiple possible counterparts ($\sum_k P_{k,j} > 0.8$ and
$P_{i,j} / \sum_{k \ne i}P_{k,j} \leq 4$) all optical objects with
$P_{k,j} > 0.15$ were included as possible matches. This resulted in
17 sources with two possible counterparts and 3 sources with three
options.

\subsection{Matching with the 24-micron data}\label{SpitzerSection}

The Spitzer 24 micron catalogue of 1194 sources in the X-ray field of
view was used to improve the X-ray to optical matching. These
data are useful as AGN often have infra-red emission, but also in a
purely statistical sense there are far fewer Spitzer sources than
optical objects in the field of view, so the probability of a random
association is far lower.  As the 24$\mu$m
sources have smaller positional errors than the X-ray sources, and are
rarer than the optical objects, a match between an X-ray and 24$\mu$m
source can significantly improve the accuracy of the optical match.

A simple likelihood ratio test, following the method for the optical
catalogue, produced 81 unique matches between the X-ray and 24$\mu$m
sources, and 20 X-ray sources with more than one possible 24$\mu$m
counterpart. The same probability cuts as the X-ray matching were
applied, and the low surface density of 24$\mu$m and X-ray sources
mean that very few false matches are expected. Examination of the
optical images for the few X-ray sources with more than one possible
24$\mu$m counterpart shows that in most cases the 24$\mu$m `point
sources' are likely to be multiple detections of nearby galaxies, which
are significantly larger than the 24$\mu$m PSF.

The 24$\mu$m sources were then matched with the optical catalogue. The
errors for the 24$\mu$m catalogue were taken as $1.2\arcsec$, which is
half a pixel. This will underestimate the errors on faint 24$\mu$m
sources, and give a more conservative catalogue. The advantage of
taking the same errors for all sources is a reduction in computing
time as only one expected distribution needs to be calculated. In
addition, as the sample was large enough, a true reliability was
calculated by comparing the likelihood ratio distribution for the
sample ($N_{true}(LR)$) with that of 10 random catalogues
($N_{random}(LR)$) following the method of \cite{Emma}. The
reliability is defined as a function of likelihood ratio,

\begin{equation}
R(LR) = \frac{N_{true}(LR)-N_{random}(LR)}{N_{true}(LR)}. 
\end{equation}

\noindent Probabilities for each possible match were calculated using Equations
\ref{prob1} and \ref{prob2}, and the criteria for unique and multiple
matches used in the X-ray matching were applied. The 24$\mu$m sources that
matched the X-ray catalogue were examined by eye to identify those which
had good optical matches, but were rejected due to underestimated
positional errors in the 24$\mu$m data.

The combined probability of the X-ray -- 24$\mu$m -- optical match was
used to identify the 24$\mu$m counterparts and to help identify
optical matches in previously ambiguous cases. For 97 X-ray sources
the addition of the 24$\mu$m data confirmed the result of the previous
matching, including 49 cases where neither the 24$\mu$m data nor the
optical catalogue gave a good match. This includes cases where the
24$\mu$m data confirm the existence of two possible matches, and cases
where two 24$\mu$m sources correspond well to one X-ray and one
optical object (and appear to be multiple detections of nearby
galaxies). For a further 14 X-ray sources the 24$\mu$m matching
identifies at least one of the optical counterparts but at a lower
probability (between 0.65 and 0.8). Four X-ray sources had a secure
24$\mu$m match with no optical counterpart, and 20 X-ray sources had
an optical counterpart but no likely 24$\mu$m match. For four X-ray
sources the 24$\mu$m data changed the assigned optical match, by
eliminating or confirming possible optical objects, and these sources
are flagged in Table \ref{A901table}.

The final list of X-ray sources with unique and multiple counterparts
is given in Table \ref{A901table}, which lists the X-ray IDs,
positions, positional error and count rates, and the possible COMBO-17
matches, optical position and photometric redshift. The probabilities and
reliabilities of the optical matches are given to allow an evaluation
of the accuracy and uniqueness of each match. The 24$\mu$m flux of any
matching Spitzer sources and the combined probability of the X-ray --
Spitzer match and the Spitzer -- Optical match both being true are
also given.

\subsection{Criteria for identifying supercluster AGN}\label{A901_AGN}

To identify the AGN in the supercluster it is necessary to use the
photometric and, if possible, spectroscopic redshifts from the
COMBO-17 survey. A cut of $0.155<z_{\rm phot}<0.185$ was used to
ensure that all AGN associated with the supercluster were identified
(see WGM05 for details of the redshift cut). This range also
allows for the errors in the photometric redshifts, which may also be
affected by the AGN emission. It is found that adding / subtracting
the COMBO-17 redshift error for each X-ray emitting source does not
reveal any more possible supercluster X-ray sources.  In addition some
galaxies have bimodal photometric redshift distributions, so the
second choice redshifts were checked and no extra supercluster X-ray
sources were found.

The presence of an AGN may cause the template fitting in the COMBO-17
survey to give wrong photometric redshifts, as the COMBO-17 templates
do not include Seyfert-like spectra with both AGN and host galaxy
contributions. To check for missed supercluster X-ray sources we
examined all optical counterparts with $21>m_{\rm R}>17.75$ (between
the faintest supercluster X-ray source and the brightest supercluster galaxy),
and $B-R<2.3$ (on or bluer than the supercluster red sequence) which
were classified as galaxies according to their template spectra.\footnote{Photometric redshifts of objects
classed as high redshift quasars are accurate as the chance of a
galaxy at $m_{\rm R}<24$ being mistaken for a quasar is very small
\citep{Wolf}.}. 
The photometric data for these galaxies were manually
compared to spectral templates at the supercluster redshift. Two
optical counterparts (COMBO catalogue numbers 12953 and 41435,
matching X-ray sources \#3 and \#135) were found to fit well with
templates at $z\sim0.16$ despite having different photometric
redshifts in the COMBO catalogue. These are discussed in detail in
the Appendix.

X-ray emission of the luminosities seen in this sample could be caused
by a large population of low mass X-ray binaries (LMXBs), hot coronae
of massive galaxies, or high levels of star formation, as well as AGN
activity. Emission from LMXBs is ruled out following the method of
\cite{Martini2}, who compare their observations to the tight relation
between B-band galaxy luminosity and the total X-ray broad band
luminosity from all LMXBs \citep{LMXBs}. Even without correcting for
the wider X-ray band used in this relation ($0.3-8$keV compared to
$0.5-7.5$keV for the A901/2 sources) the X-ray emission from the
possible supercluster X-ray sources is at least a factor of 6, and
median factor of 32 higher than the \citeauthor{LMXBs} average
relation. This is significantly higher than the scatter in their
observations. Emission from hot coronae is also highly unlikely to be
the cause of the X-ray emission, as the upper limit on the
relationship between B-band and X-ray luminosity for such sources is
very similar to that for LMXBs \citep{Sun} and the A901/2 sources have
far higher X-ray to B-band luminosity ratios.

To distinguish between X-ray emission from high levels of star formation
and that from AGN, various methods were used depending on the information
available for each source. Used alone most of these methods cannot
distinguish absolutely between star-forming galaxies and those with AGN,
but combining the available data can give a reliable indicator of the
nature of the X-ray emission.

\begin{itemize}
\item{{\bf X-ray Luminosity -- } Star forming galaxies generally have
low X-ray luminosities. A source in the local universe with $L_{0.5-8
{\rm keV}} > 3 \times 10^{42} {\rm erg s^{-1}}$ is extremely unlikely be
purely star forming \citep{Bauer}, and any source with $L_{0.5-8 {\rm
keV}} \gtrsim 1 \times 10^{41}{\rm erg s^{-1}}$ is likely to be an AGN
(see Figure 7 of \citeauthor{Bauer} -- most sources with $L_X >
10^{41}$ which are not near the flux limit are securely identified as
AGN by their column density or hardness ratio). Luminosities were
calculated using aperture photometry on images with the mean
background subtracted. The background subtraction process followed the
method of \cite{Arnaud}, and will be described in the forthcoming
paper on the extended emission (Gray et al. in prep.). The redshift of
the supercluster AGN ($\sim$0.17) means that the observed 0.5--7.5 keV
counts can be converted into an emitted 0.58--8.7 keV luminosity. As
this study does not require very accurate luminosities, and as the
errors on the count rates are large, this is taken as an approximation
to the 0.5--8 keV luminosity. The true luminosities may be slightly
higher, but this depends on the X-ray spectrum. }
\item{{\bf [OII] Star-Formation Rates -- } If the X-ray emission is purely
due to star-formation, with no AGN present, then the star-formation rate
(SFR) can be estimated from the soft band X-ray luminosity as
\begin{equation}
SFR ({\rm M}_{\odot}/{\rm yr})= 2.2 \times 10^{-47} L_{0.5-2 {\rm keV}}
({\rm W})
\end{equation}
\noindent \citep{Ranalli}. If there is no AGN and minimal absorption then
this must be near to the SFR derived from the [OII]$\lambda$3727 line flux
\citep{Hopkins};
\begin{equation}
SFR ({\rm M}_{\odot}/{\rm yr})= \frac{L_{[OII]}}{2.97 \times 10^{33} W}
\end{equation}
\noindent where $L_{[OII]}$ can be estimated for those objects with
2dF spectra from the equivalent width of the line and the COMBO-17
magnitude in the rest-frame Johnson U band. (This method assumes a
flat spectrum in the U band, but will give an estimate of the flux at
3727\AA \hspace{0.05cm} to within at least a factor of two, as the U
band magnitude has minimal contamination from flux above the 4000\AA
\hspace{0.05cm} break).  }\item{{\bf X-ray / Optical Flux Ratio -- } Sources with $f_{0.5-8 {\rm
keV}}/f_R > 1 $ are very likely to be AGN, and those with $f_{0.5-8 {\rm
keV}}/f_R > 0.1 $ are likely to be AGN (see \citealt{Bauer} and references
therein). }
\item{{\bf X-ray Hardness Ratio -- } A good indication of the spectral
properties and absorption of X-ray sources is given by the luminosity
hardness ratio, $HR=\frac{H-S}{H+S}$, where $H=L_{2-8 {\rm keV}}$ and
$S=L_{0.5-2 {\rm keV}}$. Sources with $HR>0.8$ are unlikely be
star-forming due to the very large amounts of absorption required
\citep{Mainieri} (unless the emission is dominated by hard X-ray
binaries), and sources with $HR>-0.2$ are more likely to be AGN than
star-forming \citep{Szokoly}. Hardness ratios were calculated from the
background subtracted images which will be described in detail in the
paper on extended emission in this field.}
\item{{\bf Optical Line Ratios -- } Line ratios in optical spectra can
distinguish between star-forming galaxies and AGN. As the 2dF spectra
are not flux calibrated only the [OIII] and H$\beta$ lines were used,
as they are close together in wavelength so considering the line
equivalent widths, and assuming a flat continuum spectrum, will not
introduce overly large errors. Most of the 2dF spectra of the optical
counterparts have very faint or no lines so only upper limits can be
measured. \cite{Lamareille} compare the classification of emission
line galaxies using the traditional \cite{Baldwin} diagnostic method,
and a method based on only blue emission lines ($\lambda >
3700$). Their results show that if ${\rm [OIII]}\lambda 5007/{\rm
H}\beta \gtrsim 5.5$ then the object is almost certainly an
AGN. However if the ratio is $< 5.5$ the object could still contain
some AGN activity, and if no lines are visible it may be an obscured
AGN, so this method can confirm the presence of an AGN but cannot rule
out any AGN activity. }

\end{itemize}

\subsection{Details of supercluster AGN candidates}\label{A901_AGN_det}

After applying the redshift cut, the candidates for supercluster AGN
are reduced to eleven optical matches with secure photometric or
spectroscopic supercluster redshifts, two matches with revised
photometric redshifts and three less secure matches with confirmed
supercluster galaxies. Eleven of the candidate supercluster AGN host
galaxies were observed with 2dF, and five do not have spectra.

The details of the candidates are given in the Appendix and
in Table \ref{AGN_table}, including their X-ray and optical
properties, images of the host galaxies and spectra where
applicable. The spectral energy distribution (SED) of the supercluster
galaxies is given in each case as either `old red', `dusty
star-forming' or `blue-cloud', as defined in the WGM05 sample from
photometric fits to galaxy templates.  The morphologies of the host
galaxies which are included in the sample of \cite{Lane} are also
given.

The candidates were assigned to be supercluster AGN, possible
supercluster AGN, or rejected outright. The supercluster AGN are
sources \#20, \#24, \#34, \#37, \#71, \#79, \#81, \#104, \#105, \#135
and \#139. The possible AGN is \#3. The final sample therefore
consists of 11 likely supercluster AGN and 1 possible member. Two
sources, \#3 and \#135, show significant variability in the broad band
photometry. These are also the two most luminous sources in the X-ray,
and due to the rapid variability must be optical Type I AGN.

All of the AGN hosts appear to be massive galaxies and some appear to
be morphologically disturbed. The morphologies assigned to the AGN
hosts are indistinguishable from the parent population of luminous
cluster members. Detailed study of the smaller scale morphologies of
the AGN hosts is beyond the scope of this work, and will be left until
recently obtained Hubble Space Telescope images of this field are
analysed. The fraction of `old red', `dusty star-forming' and
`blue-cloud' host galaxies (as defined in the WGM05 sample from
photometric fits to galaxy templates, and excluding the very bright
AGN as it is significantly contaminated with AGN optical light) are
0.5, 0.2 and 0.3, compared to fractions in the intermediate density
galaxy population of $\sim 0.54$, $\sim 0.26$ and $\sim 0.20$, so the
AGN hosts are again indistinguishable from the parent population.

It is worth noting that all of the supercluster AGN candidates are
classed as galaxies in the COMBO-17 survey as the photometric method
is not sensitive to low luminosity Seyfert-like AGN and obviously
misses optically obscured AGN.  In addition, of the 8 supercluster AGN
with optical spectra only 5 have emission lines, and only two have
[OIII]/H$\beta$  ratios which would lead to a classification as an
AGN using optical data alone. This highlights again the need for X-ray
studies to investigate the AGN population.

\section{Analysis of AGN properties and environments}\label{A901-analysis}

The supercluster A901/2 is very diverse and contains a wide range of
environments which may have an effect on AGN activity. Eleven X-ray
sources in this field are likely to be supercluster AGN. Using the
COMBO-17 dataset the properties of the host galaxies of these AGN can
be found, and the number of galaxies hosting AGN can be
calculated. This information can be used to construct a control sample
of galaxies which appear similar to AGN hosts, but do not have
significant X-ray emission. By looking at nearby galaxy distributions,
the environments of the AGN hosts can then be compared to those of the
control sample to determine whether environment and AGN activity are
linked.

\subsection{Properties of the AGN hosts}\label{A901_hosts}

The WGM05 supercluster sample contains 795
galaxies, where all galaxies with $0.155<z_{\rm phot}<0.185$ and
absolute Johnson V magnitude $< -17$ were identified as supercluster
members. The large redshift range (the same as applied for the AGN in
Section \ref{A901_AGN}) allows for the errors in the photometric
redshifts, and the magnitude cut removes faint objects, which have far
less accurate photometric redshifts.

\begin{figure}
\begin{center}
\includegraphics[clip,width=\columnwidth,angle=0]{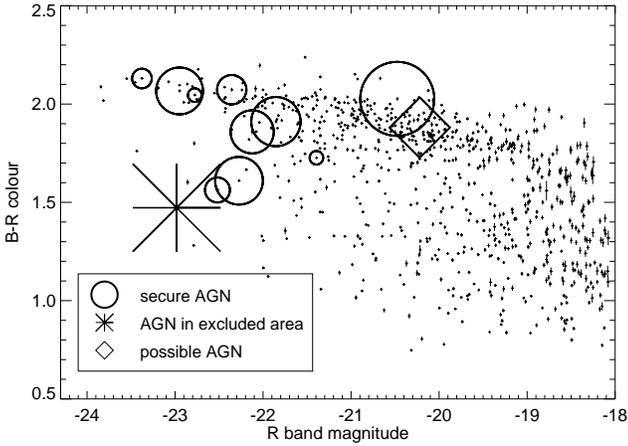}
\caption[Colour--Magnitude plot for supercluster galaxies and AGN
hosts]{The colour--magnitude plot for all supercluster galaxies (small
dots with error bars) and AGN hosts. Total absolute magnitudes are
plotted, with errors in the colour and flux measurement. The errors do
not include uncertainties in the redshifts, which increase the error
bars to $\pm 0.1$. The area of the AGN symbols is proportional to
log($L_X$(0.5--8keV)). The X-ray source \#135 is marked with a star as
it is not in the control area described in Section \ref{compare_gal}.
\label{A901colour_mag}}
\end{center}
\end{figure}

The colour--magnitude diagram for the supercluster is shown in Figure
\ref{A901colour_mag}, with the supercluster AGN host galaxies
indicated (it is important here to note that the R band magnitudes of
the host galaxies are not significantly changed by the presence of an
AGN, as can be seen directly from the 2dF spectra).  All of the AGN
and possible AGN lie in bright ($M_{\rm R}<-20$) galaxies. The X-ray
luminosity of the AGN, as indicated by the size of the symbols in
Figure \ref{A901colour_mag}, shows that the lack of AGN in fainter
galaxies is not due to fainter X-ray AGN being found in optically
fainter galaxies, which would lead to AGN in lower luminosity galaxies
falling below the detection threshold. In fact, if any correlation
exists it is likely to be the opposite -- there is an 84 per cent
chance that more X-ray luminous AGN are found in optically fainter
galaxies, according to a Spearman rank test.

\begin{figure}
  \begin{center}
\includegraphics[clip,width=\columnwidth]{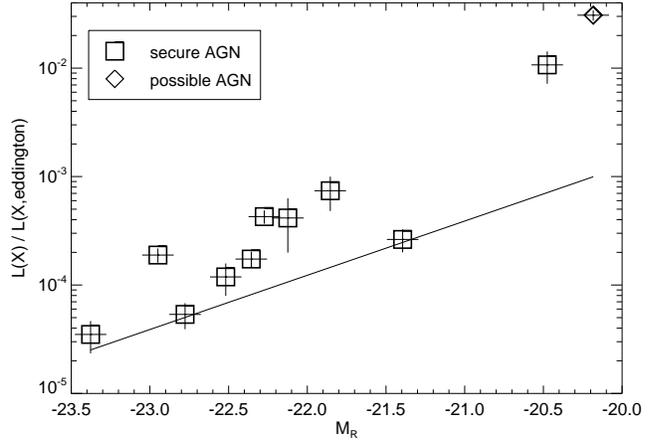}
    \caption[AGN accretion efficiency as a function of host
    luminosity.]{Approximate accretion efficiency (fraction of
    Eddington X-ray luminosity) as a function of host galaxy absolute
    magnitude. The solid line shows the approximate minimum detectable
    accretion efficiency at each point. The X-ray bright source \#135
    is excluded from this plot, and has an accretion efficiency of
    $\sim 4$ per cent at $M_{\rm R} = -23$.}
    \label{A901_agn_mra}
    \end{center}
\end{figure}

To find the range of accretion rates covered by this sample, $L_{\rm
X}/L_{\rm X,eddington}$ was calculated for each AGN, where it was
assumed that 10 per cent of the bolometric luminosity is emitted in the
0.5--8keV band.  The relation $log(M_{BH}/M_{\odot}) = -0.5 M_{\rm R} -
2.96$ of \cite{McLure} was used to calculate the black hole mass from
the rest frame R-band absolute magnitude ($M_{\rm R}$) given in the
COMBO-17 catalogue (as derived from the galaxy template, corrected to
the cosmology of this paper). The resulting plot (Figure
\ref{A901_agn_mra}, excluding the bright source \#135), although crude
due to the approximations made, appears at first glance to show a
correlation between $M_{\rm R}$ and $L_X/L_{X,eddington}$ (which would
not be changed by altering many of the assumptions above, for example
the 10 per cent emitted in the X-ray, as they would only change the scale
of the plot). However, there are more galaxies at $M_{\rm R} \sim -21$
than $M_{\rm R} \sim -23$ and for the fainter galaxies lower efficiency
AGN may fall below the detection limit. To calculate whether the lack
of more efficient AGN in more luminous galaxies is due to the smaller
sample size or a physical effect it is necessary to know the number of
possible AGN hosts as a function of $M_{\rm R}$.

\begin{figure}
\begin{center}
\includegraphics[clip,width=\columnwidth]{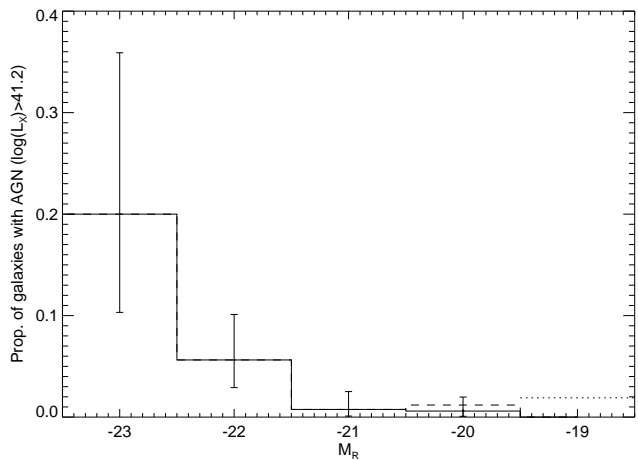}
    \caption[Proportion of galaxies hosting AGN as a function of host
    magnitude.]{Proportion of galaxies hosting AGN as a function of
    R-band absolute magnitude, with $1\sigma$ error bars. The dashed
    lines show the results if the possible supercluster AGN is
    included. The dotted line in the final bin shows the 95 per cent
    confidence limit for the first empty bin. Again source \#135 and
    galaxies where AGN activity could not be easily detected are not
    included.}
    \label{A901_agn_mrb}
    \end{center}
\end{figure}

\subsection{The fraction of galaxies containing a detected AGN}\label{compare_gal}
To determine the proportion of galaxies with AGN and the properties of
the AGN hosts it is necessary to define a control area in which AGN
could have been detected, and compare the galaxies with and without an
AGN in this area.  To select the control area the COMBO-17 catalogue
was cut to remove objects within $160\arcsec$ of the top and sides and
$300\arcsec$ of the bottom of the image. This cut ensures the returned
area is 97 per cent covered by the X-ray image, and also ensures that
the edge of the catalogue does not affect properties such as local
density. In addition areas where the sensitivity to point sources
decreased significantly due to extended sources or very bright X-ray
emission, were removed from the control area: three areas were bright
enough to obscure moderate luminosity AGN: a $30\arcsec$ circle around
the extended emission to the north-west of A901a (marked with a
rectangle in Figure \ref{xray_sources}), $67\arcsec$ around the AGN in
A901a (\#135) and $83\arcsec$ around A901b. In these regions the noise
level is at least 50 per cent higher than for the rest of the
field. At the edge of the excluded regions the faintest detected
supercluster AGN (1.8$\times 10^{42}$erg/sec) would be detectable at
low significance, but the sensitivity towards the centre of \#135 and
A901b decreases to a level where only AGN brighter than $\sim
10^{44}$erg/sec would be detectable. No sources are detected in these
areas (except for \#135 itself), even at low significance. The small
changes in detection sensitivity due to emission from A902 and the SW
group and the changes in PSF were not included as they only affect
very marginal detections, such that the faintest supercluster AGN
would still be detected at low significance, and a small change in
sensitivity is not significant in a sample of this size. The applied
cuts also remove the AGN in A901a (\#135) from the sample, which is
necessary as it obscures possible AGN activity from all of the
surrounding large galaxies so will bias the sample. In addition its
high accretion efficiency and luminosity and optical variability show
that it, and the one possible supercluster AGN, are the only X-ray
Type-I AGN in the sample. The control sample contains 604 supercluster
galaxies, as 149 were removed in the edge cut and 42 were in regions
where AGN could not be detected.

The number of AGN hosts in the supercluster can be directly compared
to the number of possible host galaxies in the control sample area.
The number of AGN per possible host galaxy is shown in Figure
\ref{A901_agn_mrb} for a range of host luminosities.  The total
fraction of galaxies with $M_{\rm R}<-20$ hosting AGN with $L_X >
10^{41.2}$erg s$^{-1}$ is 10/253, or $\sim$4 per cent. This is similar
to the results of \citet{Martini2}, who found a $\sim$5 per cent X-ray
detected AGN fraction for galaxies with $M_{\rm R} < -20$ in low
redshift clusters at a flux limit of $L_X = 10^{41}$erg
s$^{-1}$. Figure \ref{A901_agn_mrb} also shows an increase in AGN
fraction (above the luminosity limit) in increasingly massive
galaxies, but the small sample size means that the error bars are very
large.  In this supercluster AGN are only found in galaxies with
$M_{\rm R} < -20.4$.  As stated previously, the luminosities of the AGN
show that this is not due to the flux limit of the sample. Rather, it
appears that fainter AGN are more likely to be found in more luminous
galaxies, brighter AGN in moderately luminous galaxies, and no AGN in
the least luminous galaxies.

Returning to Figure \ref{A901_agn_mra}, it is now clear that the lack
of AGN with $L_X/L_{X,eddington} > \sim 3 \times 10^{-4}$ and $M_{\rm
R} < -22.5$ is probably due to the lack of bright galaxies, rather than a
tendency for more luminous host galaxies to have lower efficiencies:
there are only 19 possible AGN host galaxies above this luminosity,
and as $\sim$3 per cent of fainter ($-22.5 < M_{\rm R} < -21.5$) galaxies
have accretion efficiencies above $3 \times 10^{-4}$ it is not
surprising to find no bright galaxies with AGN efficiencies above this
level. Similarly, the one supercluster AGN with efficiency $> 10^{-3}$
only corresponds to $\sim$0.5 per cent of the galaxies in that 0.5
magnitude bin, explaining the lack of brighter galaxies with similar
efficiencies.  However there must be a significant decrease in the
efficiency of any AGN in galaxies with $M_{\rm R} > -20 $ as the
number of supercluster galaxies is very large yet no AGN are observed
above the X-ray flux limit.

\subsection{Defining a control sample}\label{control_samples}

To compare the AGN environments and properties, control samples were
created, consisting of galaxies similar to the AGN hosts, where AGN
activity could have been detected but was not found. Whereas a
randomly selected control sample would contain many faint galaxies, it
is instead preferable to define a control sample with a similar
distribution of galaxy properties as the AGN hosts. Any difference
between the AGN hosts' environments and the control sample
environments would therefore be due to an environmental effect on AGN.

100 control samples were made, each consisting of 65 of the 183
supercluster galaxies with aperture magnitude $m_{\rm ap,R} < 20$
which lie in the control area described in Section \ref{compare_gal}
and are not AGN hosts. These samples were selected at random such that
there are equal numbers of galaxies in each 0.5 aperture magnitude bin
to replicate the distribution of AGN host magnitudes. This method
ignores the possible increase in the number of AGN in brighter
galaxies (Figure \ref{A901colour_mag}) but this is less significant when
aperture magnitudes are used, and due to the small number of galaxies
selected should not affect any results significantly.

The 100 samples are identical at $m_{\rm ap,R}<18.5$, due to the small
number of available galaxies, but at $18.5<m_{\rm ap,R}<20$ different
sets of galaxies were chosen, although the samples still have
considerable overlap with each other. Each of the 100 control samples
was compared to the AGN sample and the median statistic taken to
reduce the errors.  A Kolmogorov--Smirnov (K--S) test (to identify
changes in the mean) and Kuiper test (a K--S test using Kuiper's
statistic which is better at identifying changes in spread) confirm
that the control samples and AGN hosts are drawn from the same R-band
magnitude distribution (at at least 56 per cent confidence for the supercluster
AGN).

The colour--magnitude diagram in Figure \ref{A901colour_mag} also
shows that at least three AGN hosts are significantly bluer than the
red sequence.  K--S and Kuiper's tests on the deviation from the red
sequence give 0.59 and 0.48 probabilities that the supercluster AGN
hosts and control galaxies are drawn from the same
distribution. \cite{Martini} found a propensity for AGN hosts to be
bluer than similar galaxies in the cluster A2104, but this sample is
so small that such an effect cannot be confirmed here.

\subsection{The local galaxy densities of AGN host galaxies}

\begin{figure}
\begin{center}
\includegraphics[clip,width=\columnwidth,angle=0]{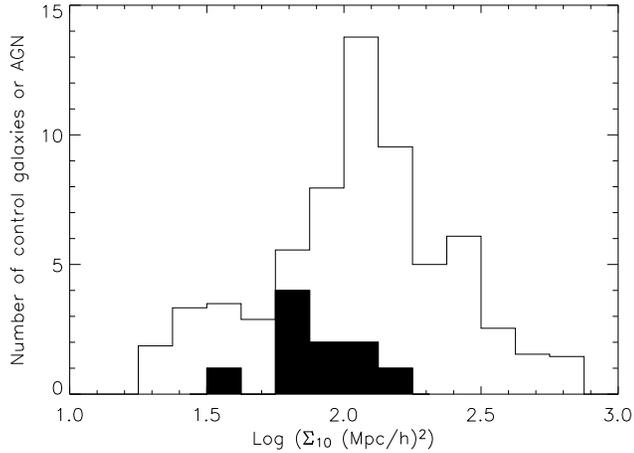}
\caption[]{Distribution of $\Sigma_{10}$ for the control samples
(solid line, mean value of all 100 samples is given) and supercluster
AGN host galaxies (filled histogram). The control sample does not
include galaxies in three areas of high X-ray emission, where moderate
luminosity AGN could not be detected. The AGN host galaxies have far
lower values of $\Sigma_{10}$, and a narrower spread of values, than
other similar galaxies. The difference between the control samples and
AGN (taking the median value) is significant to $>0.98$ using Kuipers
test and $>0.99$ using a K--S test.}\label{sig10}
\end{center}
\end{figure}

The environments of the AGN were evaluated by comparing the local
densities of the AGN host galaxies to those of the control
samples. The surface mass density from the weak-lensing analysis in
\cite{Gray02} was not used as the sample of AGN is small and they lie
in areas (outside cluster cores) where the errors are quite large,
such that any results would have very low significance. Instead, the
projected galaxy density was used, following the method for the WGM05
sample, where $\Sigma_{10}$ is defined as the number of supercluster
galaxies per (Mpc h$^{-1}$)$^2$ within a circle with a radius given by
the average distance to the 9th and 10th nearest neighbours. The
supercluster sample uses only the brightest galaxies as defined in
Section \ref{A901_hosts}.
 
The distributions of $\Sigma_{10}$ for the control samples and the
supercluster AGN are shown in Figure \ref{sig10}. The AGN host
galaxies lie predominantly in areas of moderate density compared to
the control sample, with a significance of 98.4 per cent and 99.0 per
cent from Kuipers and the K--S test respectively. It is particularly
clear that AGN hosts avoid the densest regions of the supercluster,
even accounting for the fact that two of the cluster cores are removed
from this study. The exception, of course, is the very luminous AGN in
the core of A901a, which was excluded, along with the surrounding
galaxies, as they do not lie in the control area. This source is
different in both properties and position from the other AGN in the
supercluster.

\subsection{The types of environments that contain AGN}

\begin{figure}
\begin{center}
\includegraphics[clip,width=\columnwidth,angle=0]{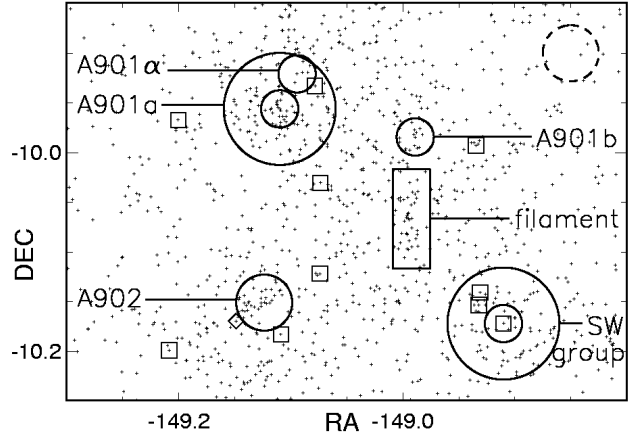}
\caption[Defined regions within A901/2 and the positions of the
AGN]{Regions with a clear environmental category in the
supercluster. Regions are identified by eye, using deep images and
galaxy properties as well as the galaxy positions shown. Galaxies in
the 80 per cent complete supercluster sample are marked with a small
cross, and for comparison the AGN are also marked by squares
(supercluster AGN) and diamonds (possible supercluster AGN). The
figure is around $30 \arcmin$ by $24 \arcmin$, with north-east to the
top left. The dashed circle in the top right-hand corner is $1.5
\arcmin$ in radius, the size used to determine the local environmental
properties of each galaxy.\label{A901_environmentareas}}
\end{center} 
\end{figure}

The supercluster contains a wide range of environments which are
evident by eye, such as clusters, groups and filaments of galaxies. In
order to better classify the environments of AGN, it is desirable to
find a method of distinguishing between these environments in terms of
their properties. As a first step, areas which fall into a clear
environmental category were selected by eye, as shown in Figure
\ref{A901_environmentareas}.  Although arbitrarily defined, the
properties of galaxies in these areas can help to work out what
properties distinguish different environments, and hence to define the
environments of other supercluster galaxies and the AGN hosts.

The regions were defined as follows: cluster and group regions are
circles of $1\arcmin$ radius, centred on the brightest cluster
galaxy. As A902 does not have a clear centre, a radius of $1.5\arcmin$
is used to include the whole cluster region. For the rich cluster
A901a and the SW group a second region out to $3\arcmin$ radius is
considered, marking the cluster and group edge environments. The
filament region, with a large number of blue galaxies, is marked with
a rectangle.

\begin{figure}
\begin{center}
\includegraphics[clip,width=\columnwidth,angle=0]{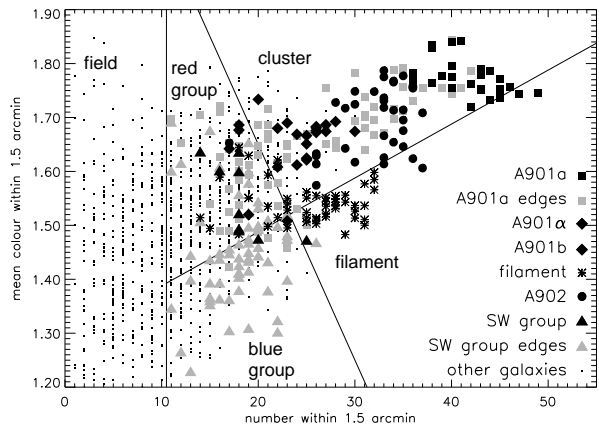}
\caption[Local density vs. Local colour plot for galaxies in well
defined environments]{The local number density and mean local B-R
colour for galaxies in the 80 per cent complete supercluster
sample. Galaxies in the defined environments of Figure
\ref{A901_environmentareas} are marked with larger symbols. Galaxies
in A901a and the SW group are split into inner and outer
regions. Typical uncertainties are $\pm$5 in local number and
$\pm$0.075 in mean colour assuming a random distribution. The true
variation in mean colour is generally lower as different regions of
the supercluster are dominated by different colour galaxies. The
plot is divided into five regions, containing galaxies in different
types of environment -- for example the `cluster' region contains
mainly galaxies in cluster cores and the outskirts of large
clusters. The definition of the regions is justified in the text.
\label{A901_gal_colnum}}
\end{center}
\end{figure}

 There are many possible ways to define galaxy environments, such as
mean local luminosity, galaxy colour or distance from the rest frame
U-V colour-magnitude main sequence (as used by \citealt{Gray}), but
not all of these were found to be useful in distinguishing between the
predefined areas in Figure \ref{A901_environmentareas}.  It was found
that the galaxies in different environments could not be separated
well purely in terms of local projected density, and that two other
factors are also important: firstly the number of less luminous
galaxies, and secondly the colour of the local galaxies. The distance
from the colour-magnitude relation was investigated as a measure of
colour, but the environments were better separated using the measured
colour, as the latter combines colour with a distinction between
bright and faint galaxies. The mean local colour was used as this
measure is sensitive to the presence of a few very blue or active
galaxies, wheras the median colour is dominated by red galaxies on the
colour-magnitude main sequence (Figure \ref{A901colour_mag}), and is
more a measure of local magnitude. To include fainter galaxies the
supercluster sample was widened to include galaxies with aperture
magnitudes $m_{\rm R} < 23$. At this magnitude the redshift errors are
larger, and so a magnitude dependent redshift cut was used to give an
80 per cent complete sample. The magnitude dependent redshift cut
traces the increase in photometric redshift errors, and is given by
\begin{equation}
0.17-\sigma(m_{\rm R}) < z_{\rm phot} < 0.17+\sigma(m_{\rm R})
\end{equation}
where
\begin{equation}
\sigma(m_{\rm R}) = 0.0075 \times \sqrt{1+10^{0.6(m_{\rm R}-20.5)}}
\end{equation}
with a minimum cut of $\sigma(m_{\rm R}) = \pm$0.015 as used
previously. The broader redshift cut starts to take effect at $m_{\rm
R} > 21.75$, and the level of contamination is $\sim 20$ per cent at $m_{\rm
R}=22$ and $\sim 40$ per cent at $m_{\rm R}=23$.
Although the broader sample includes far more contamination from field
galaxies, this will be uniform across the field of view. Plotting the
spatial positions of the fainter galaxies shows that a significant
number are associated with the supercluster, and these are
instrumental in distinguishing between the different environments in
the following analysis.

Using this sample, the number and mean B-R colour of all supercluster galaxies
within a projected distance of 1.5 arcmin was found for each
supercluster member, which corresponds to 250kpc at
the supercluster redshift. The size of this region is illustrated in
Figure \ref{A901_environmentareas}, and was chosen to be large enough
to include a statistically significant number of galaxies, with an
average of 14 supercluster members within this radius for objects in
the control area.

Plotting the local galaxy density and local mean colour is a good
 method of distinguishing between different environments, as shown in
 Figure \ref{A901_gal_colnum}.  The galaxies in clusters all lie on a
 local number density and local colour `main sequence' of increasing
 local number and red colour, whereas the environment of filament
 galaxies is considerably bluer than that of cluster galaxies at the
 same number density. It is conceivable that the slope of the `main
 sequence' could be caused by contamination by field galaxies, as
 galaxies with higher redshift errors are fainter and therefore bluer
 than secure supercluster members.  Although these field galaxies will
 be uniformly distributed across the field they will have a larger
 effect on the mean local colour in sparsely populated areas. Indeed
 the faint galaxies are found to be a major contributor towards the
 slope of the `main sequence'. However a closer analysis finds that
 the faint galaxies which contribute to the local colour and density
 at the dense red end of the `main sequence' are significantly
 ($\sim$0.2 magnitudes) redder than those contributing to the group
 end. This fact, together with their spatial distribution, confirms
 that significant numbers of the faint galaxies are supercluster
 members, and shows that the colour variation between different
 environments is not caused by random field contamination but rather
 is influenced by changes in the faint galaxy population.

 Locations in the local colour -- number plane can now be used as
 a means of defining the local environment for other supercluster
 galaxies, which don't lie in the defined regions in Figure
 \ref{A901_environmentareas}. The regions are separated as shown in
 Figure \ref{A901_gal_colnum}: a line parallel to the best fit to the
 `main sequence' is used to mark the boundary between cluster and
 filament-like environments, and a line perpendicular to this marks
 the beginning of group-like environments. The redder part of the
 group-like environments contains galaxies in the outskirts of A901a,
 so this region is split into red group-like and blue group-like
 environments by a continuation of the cluster/filament
 boundary. Finally, the field-like environment was selected to contain
 none of the galaxies from the defined regions.

\begin{figure}
\begin{center}
\includegraphics[scale=0.5,clip]{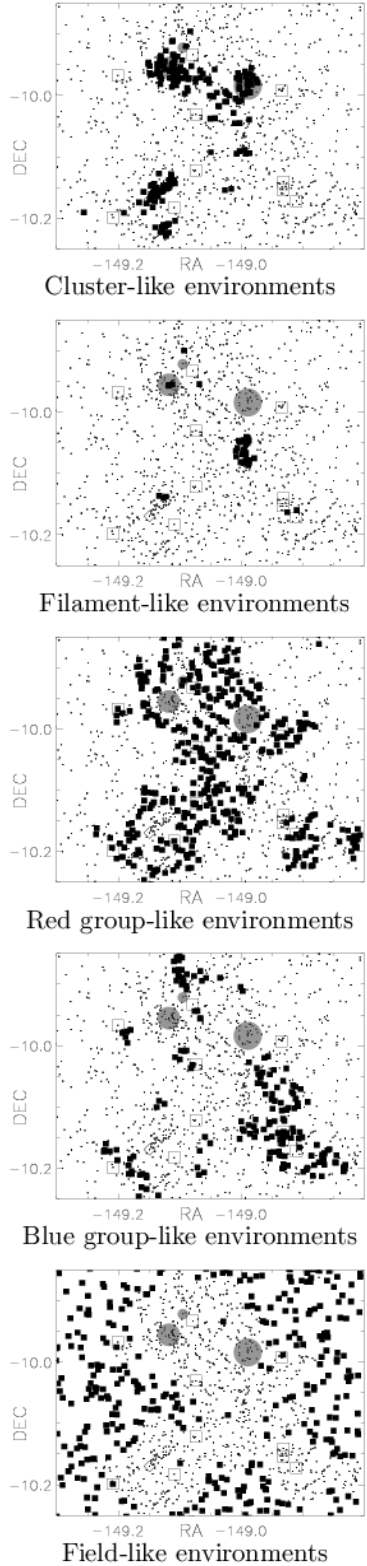}
\caption[Galaxies in the 80 per cent complete supercluster sample, separated
    according to environment.]{Galaxies in the 80 per cent complete
    supercluster sample, separated according to environment, as
    defined in Figure \ref{A901_gal_colnum}. The filled squares show
    galaxies in each type of environment, and dots show other
    supercluster galaxies. Supercluster AGN and
    possible supercluster AGN are marked with open squares and
    diamonds respectively. The grey circles mark regions where AGN
    could not be detected easily due to high levels of X-ray
    emission. Galaxies in these regions are not included in the control sample. }
    \label{A901_xy_galregions} \end{center}
\end{figure}

Figure \ref{A901_xy_galregions} shows the distribution of supercluster
galaxies in each type of environment -- it is clear that local colour
and density are indeed excellent at separating environments. This
classification by local environment rather than by arbitrarily defined
region is important -- for example, although some galaxies in the
cluster-like environment are not actually in the defined clusters, the
galaxy will experience many of the same environmental effects as a
cluster galaxy. In addition this method identifies regions such as
small groups which are difficult to define by eye. The red and blue
group-like environments contain mainly galaxies in cluster outskirts
and the SW group respectively. It is worth noting that the same
analysis for local density alone, or by $\Sigma_{10}$, is far less
successful, with mixing between filaments and cluster cores, and also
groups and smaller clusters. Also, by plotting local colour,
rather than deviation from the colour-magnitude red sequence, this
plot also has a mass dependence, as the bluer galaxies are generally
less massive.

\begin{figure}
\begin{center}
\includegraphics[clip,width=\columnwidth,angle=0]{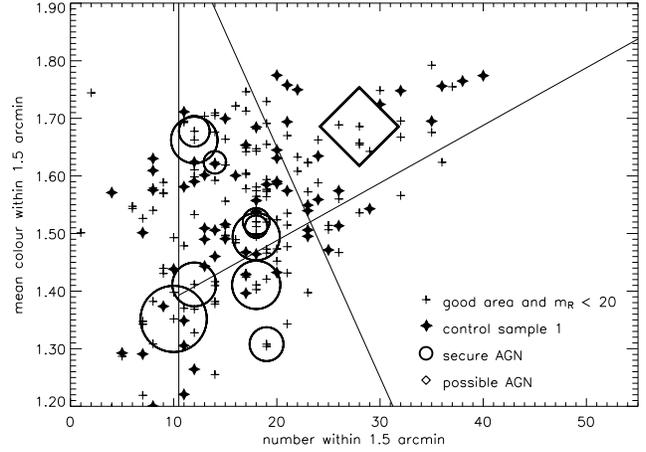}
\caption[Local number density and local mean colour for the AGN and
    control sample]{Local number density and local mean colour for the
    supercluster AGN, one possible supercluster AGN, and one of the
    control samples, with the regions defined in Figure
    \ref{A901_gal_colnum}. The crosses and filled stars mark galaxies
    in the control samples, i.e. those galaxies which are as bright as
    the AGN hosts and in regions where AGN emission could be detected
    (galaxies in the A901a/b cluster cores are not included in
    the control sample, so there are less galaxies in red, dense
    environments compared to Figure \ref{A901_gal_colnum}). A typical
    control sample of 65 galaxies with a similar distribution of
    luminosities to the AGN hosts, is indicated by the filled
    stars. Note that there are two AGN with very similar environments
    at $\sim(18,1.52)$. It is clear that the AGN have a very different
    distribution from a typical control sample. The size of the AGN
    symbols increases with increasing Log($L_X$). The excluded AGN
    \#135 would be at x=47, y=1.74. \label{A901_agn_colnum}}
\end{center}
\end{figure}

The environments of the AGN and control samples can now be evaluated
in terms of this scheme. Figure \ref{A901_agn_colnum} shows the
distribution of the supercluster AGN (excluding the bright source
\#135) and that of the control samples described in Section
\ref{control_samples}. As the control samples do not include galaxies
in three regions where AGN activity could not be detected (shown in
Figure \ref{A901_xy_galregions}) there are less galaxies in the top
right of this figure compared to Figure \ref{A901_gal_colnum}, but
there are still many control galaxies in the cluster-like
environment. Figure \ref{A901_agn_colnum} shows that the supercluster
AGN all lie within or very close to the red and blue group-like
environments in local number -- colour space. This is very different
from the distribution of the control samples, where a significant
proportion lie in filament or cluster-like environments, as well as in
the field. In particular the AGN hosts in slightly higher density
areas (15 to 20 local galaxies) are all found in bluer areas, whereas
in the control sample the average local colour at this density is far
redder. A two-dimensional K--S test in the local number vs local mean
colour plane gives a probability of only 0.042 that the supercluster
AGN were drawn from the control samples. Comparing with the control
samples in each region separately, the probability that the AGN are
drawn purely from the control sample in the red and blue group-like
environments is 0.27, compared to $<0.001$ for the other environments.
In addition a Kuiper's test on the deviation from the line
perpendicular to the local colour / density `main sequence' gives a
probability of 0.009 that the AGN hosts are drawn from the same
population as the control samples (compared to 0.3 for deviation in
density only). It can therefore be asserted, at a $>98$ per cent
confidence level, that the prevalence of AGN is affected by the
environment. With the exception of the very densest areas ($\gtrsim$35
galaxies within 1.5 arcminutes, where we are not generally sensitive
to X-ray emission from AGN) the supercluster AGN are preferentially
found in moderate density red or blue environments or slightly denser
environments if the local environment is also bluer.

To search for large scale or local effects, the local number and mean
colour of the AGN and control samples were also investigated taking
regions of radius 0.5\arcmin and 3.0\arcmin. There are no significant
deviations (to $\sim$95 per cent) between the AGN and control samples, so there is
no evidence for AGN activity being triggered by very local effects,
such as mergers, and also by larger scale effects.

As an aside, it is interesting to note that both the possible
supercluster AGN and the excluded source \#135 do not follow the trend
for the main sample of supercluster AGN in Figure
\ref{A901_agn_colnum}. The co-ordinates where \#135 would fall is given
in the figure caption. Both of these AGN are confirmed as optical
Type-I by their high ($> 3\sigma$) optical variability, which is not
seen in the other sources. Source \#135 is in a very different
environment from the rest of the AGN, being very close to the centre
of the A901a cluster, but the possible supercluster AGN host, \#3, is
in a less distinct environment.

\section{Conclusions}\label{A901-conclusions}

\subsection{AGN host galaxies and properties}

This sample consists of 11 X-ray sources which are likely to be AGN in
the supercluster A901/2, with $L_{X, (0.5-8 keV)} \gtrsim
10^{41.2}$erg s$^{-1}$. Ten of these sources lie outside the very
cores of the clusters. The one bright AGN very near the centre of
A901a is not included in this study as it is in an unusual
environment, being near the centre of a massive potential well, as
well as being in an area that has to be excluded from the analysis due
to the high level of X-ray emission, to the extent that moderate
luminosity AGN could not be detected in this region. The purpose of
this study is not to investigate the most extreme environments, but
rather to focus on more `normal' galaxies and environments, beyond
$\sim$200 kpc from the cluster centre. It is possible that a source
which is classed as a `supercluster AGN' is actually a galaxy with a
highly obscured massive starburst, or a chance alignment with a
background quasar.These scenarios are unlikely, as explained in
previous sections, and removing a source from the sample will not
change the conclusions of this study. The properties of the AGN host
galaxies and a comparison with the galaxies without AGN led to the
following results;

\begin{itemize}
\item{The AGN host galaxies are all luminous, with $M_{rm R}
< -20$. Most of the host galaxies lie on the cluster
colour--magnitude red sequence, but a number are bluer, possibly due to
emission from the AGN. Of the 8 AGN with optical spectra, only two
would have probably been identified as AGN from optical data alone.}
\item{There is a no increase in AGN luminosity with optical luminosity
of the host galaxies. Formally, the best fit correlation is negative,
with the brightest AGN lying in galaxies with $M_{\rm R} \sim -20$, but this
is of low significance.}
\item{More luminous galaxies are more likely to host
an AGN above the X-ray flux limit. However, when the AGN
accretion is measured in terms of luminosity per unit black hole mass,
the small sample size and flux limits mean that that no clear
correlation is seen.}
\item{There is no propensity for AGN host galaxies to have a
particular morphology, as defined in \cite{Lane}.}
\end{itemize}

These results are generally unsurprising. The preference for optically
detected AGN to lie in more luminous galaxies is well known (see for
example \citealt{Ho}, and \citealt{Kauffmann}), and is even more
pronounced for radio-loud AGN (e.g. \citealt{Best05}). The fraction of
galaxies hosting AGN is comparable to the \cite{Martini} result for
the cluster A2104.

The lack of direct correlation, and possible inverse correlation
between $L_X$ and $M_{\rm R}$ is not what might have been
expected. Similar results were found for cluster AGN by
\cite{Martini2}, who find that galaxies more luminous than $\sim M_{\rm
R} = -21.5$ are more likely to host faint rather than bright X-ray
sources, whereas no such statement could be made for fainter
galaxies. This is possibly due to the effect of the supercluster
environment. If the AGN are being triggered by the supercluster
environment then smaller galaxies are more likely to suffer large
gravitational disturbances which can affect the central regions of the
galaxy, and perhaps provide fuel for the AGN. However, it may simply
be that larger galaxies are older and have used up much of the gas in
the central regions in previous AGN activity.

\subsection{AGN environments}

The environments of the AGN host galaxies were compared to those for a
control sample of similar galaxies in which AGN activity could have
been detected but was not (the cores of three clusters were therefore
excluded from this analysis). The following conclusions were drawn;

\begin{itemize}
\item{The AGN host galaxies are in areas of moderate galaxy density,
and strongly avoid the densest areas, with the exception of one AGN
very near the centre of a cluster. The distribution of $\Sigma_{10}$
for AGN is different to that for the control samples with $\sim 99$
per cent significance.}
\item{There are strong correlations between AGN activity and the local
environment within 250 kpc, but much less so with larger scale or
smaller scale environments. There is no correlation with distance to
the nearest neighbour.}
\item{The environments in A901/2 can be split according to local
galaxy number density and local mean B-R colour (which also traces
mass as smaller galaxies are bluer). The distribution of the AGN host
galaxies is significantly ($\gtrsim$98 per cent) different to that of
the control sample, with AGN found in areas of moderate density or
slightly higher density and bluer colour, similar to cluster outskirts
and groups. Galaxies in cluster-like environments, the blue filament
and the field are less likely to host AGN. No secure AGN are found in
areas with $\sim 20$ nearby galaxies but red colours, such as the
centres of small clusters, whereas AGN are found in similar density
environments with more blue galaxies.}
\end{itemize}

It is clear from these results that the AGN are affected by their
environment, although the small sample size requires caution when
drawing conclusions. As three cluster centres (with radii of 90, 190
and 240 kpc, shown in Figure \ref{A901_xy_galregions}) are not studied
due to the high X-ray background, conclusions cannot be drawn for
cluster cores, but there are still many control galaxies in high
density regions that do not have AGN activity.  The lack of AGN in low
density regions suggests that the AGN activity is enhanced relative to
the field, and the lack of AGN in the highest density regions shows
that this enhancement is followed by either suppression or a return to
the field values. The main conclusion, that AGN lie predominantly in
cluster outskirts and groups, suggests that AGN are triggered by
joining a denser environment.  \cite{Heckman} find that the peak of
optically detected AGN activity ([OIII] emission) occurs in galaxies
with properties between those of young star-forming galaxies and old,
quiescent bulge-dominated galaxies. These results suggest that AGN
activity could be linked to the transformation of field galaxies to
cluster members.

This result at first appears to be in contrast to that of
\cite{Martini_pos}, who show that AGN with $L_{\rm X}>10^{41}$erg
s$^{-1}$ in eight clusters have a radial distribution, measured from the
cluster centre, which is consistent with that for the luminous
galaxies without AGN. However, when the same analysis is performed on
A901/2, using the distance to the nearest cluster (both including and
excluding the SW group), there is also no significant difference
between the distributions for the AGN and control samples. The links
between environment and AGN activity in this supercluster are
therefore not detectable if only cluster-centric distance is compared,
and are not in opposition to the results of \citeauthor{Martini_pos}.

The AGN exist in areas of moderate density, but also in slightly
denser areas which are dominated by bluer galaxies. Two possible
interpretations for this are either that the bluer mean colour
indicates a higher level of activity in the surrounding galaxies, or
that the local environment is dominated by smaller galaxies. In the
latter case, if the bluer environments of some AGN are due to a larger
number of small galaxies, this would possibly suggest minor mergers
with small galaxies as the trigger for AGN activity.  Adding a third
dimension of local mean $M_{\rm R}$ may help to resolve this issue,
but the current sample is considered too small for this analysis.

If, on the other hand, the bluer environments of some AGN are due to
areas with enhanced star-formation, it suggests that the same
processes that trigger star-formation also trigger AGN activity. This
would agree with the results of \cite{Coldwell2}, who find that
optically identified AGN lie in regions containing more emission line
galaxies. In A901 the correlations between optically detected
star-formation and density \citep{Gray} show that star-formation is
suppressed in cluster cores, rather than triggered in cluster
outskirts. In contrast the photometrically detected ``dusty
star-forming galaxies'' in the WGM05 sample are, like the AGN hosts,
mainly found in intermediate density environments. However, there is
no propensity for AGN to reside in galaxies of any
particular type in this scheme. Also no correlation is found between
the WGM05 host type and the AGN local density, luminosity or position.
The transformation processes which create the dusty red population
could also cause AGN activity, but the two populations do not appear
to be linked directly.

In conclusion, in this supercluster AGN activity is strongly linked to
environment, and occurs predominantly in moderate density regions,
which are often bluer than average for this supercluster, similar to
(but not limited to) cluster outskirts and blue groups.

\vspace{-0.4cm}
\section*{Acknowledgments}
The authors wish to thank Kyle Lane and Alfonso Arag\'{o}n-Salamanca for
supplying the morphologies of the AGN hosts from their sample in
\cite{Lane}. OA and PNB acknowledge the support of the
Royal Society. CW was supported by a PPARC Advanced Fellowship.
\vspace{-0.5cm}




\newpage
\section{Appendix}\label{appendix}

This appendix contains details of the individual X-ray sources which
were possibly supercluster AGN and their classifications. The images
and spectra, where applicable, are shown in Figures \ref{NospecImages}
and 14
, and details are given in Table \ref{AGN_table}.

\begin{itemize}

\begin{table*}
\begin{tabular}{rrrrll}
X-ray ID&L$_{\rm X}$ (erg/sec)& HR   & Optical ID& Redshift& Optical properties  \\
\hline
3  &  5.25$\times 10^{42}$& 0.18 & 12953 & 1.4(phot)   & variable, blue-cloud or contaminated by AGN \\
20 &  8.1$\times 10^{41}$ & 0.03 & 31178 & 0.158       & Sa,  blue-cloud\\
24 &  7.8$\times 10^{41}$ & -1.0 & 36966 & 0.170       & E, old-red \\
34 &  6.9$\times 10^{41}$ & 0.7  & 39549 & 0.163       & Sab, dusty-starforming\\
37 &  2.4$\times 10^{41}$ & -1.0 & 14161 & 0.171(phot) & E, old-red\\
71 &  3.0$\times 10^{41}$ & 0.53 & 11827 & 0.175(phot) & Sb, blue-cloud\\
79 &  1.8$\times 10^{41}$ & -0.75& 44351 & 0.161       & E, old-red \\
81 &  1.8$\times 10^{41}$ & -1.0 & 19305 & 0.168       & S0, blue-cloud\\
104&  8.6$\times 10^{41}$ & 0.95 & 15698 & 0.170       & Sab, dusty-starforming\\
105&  3.6$\times 10^{41}$ & 0.0  & 17675 & 0.171       & E, old-red\\
135&  1.55$\times 10^{44}$& 0.03 & 41435 & 0.33(phot)  & early-type, variable, contaminated by AGN\\
139&  2.56$\times 10^{42}$& 1.0  & 9020  & 0.169       & S0, old-red\\
\end{tabular}
\caption{Properties of supercluster AGN and possible supercluster AGN. Details of each source are given in the text.} \label{AGN_table}
\end{table*}

\item{{\bf \# 3 -- possible supercluster AGN.} The high X-ray flux
indicates that this source is clearly an AGN if it is in or behind the
supercluster. The host galaxy identification is secure, with a
photometric redshift of 1.4, but the optical object is highly variable
and the broad band photometric data points taken on different days
vary by up to 15$\sigma$. This indicates AGN activity, and
this contamination may explain the photometric redshift. It is very
unlikely that the redshift of 1.4 is correct as the galaxy would then
have to be three magnitudes more luminous than the brightest
supercluster member. However, a visual fit to a $z=0.16$ galaxy
template (Figure \ref{specfit_12953}) is fair, suggesting that
it may be in the supercluster. In addition the colours of this galaxy
place it on the cluster main-sequence. Over 77 per cent of similar
galaxies (within $\delta m_{\rm R} = 0.5, \delta {B-R} =
0.1$) are within the supercluster. }

\item{{\bf \# 11 -- rejected.} A high hardness ratio (0.49) and
moderate flux indicate a probable AGN. However, there are two possible
optical counterparts, and only object \#51604 is in the
supercluster. The spectrum and image of the supercluster member show a
very luminous Sa galaxy with an `old red' SED and no emission
lines. The non supercluster member is both closer to the X-ray source
and a probable quasar, which are often X-ray luminous ($\sim$50 per
cent of the COMBO-17 QSOs with $m_{\rm R}<23$ are detected as X-ray
sources, and $\sim 15$ per cent of the possible QSOs are also
detected) and are rare compared to galaxies.  Combining the distance
of each object from the X-ray source and the number of bright QSOs in
the field to that of bright galaxies, both matches are $>$99
per cent significant. However, the match with the QSO is four times
more probable, and when the Spitzer data is taken into account the
QSO match becomes even more likely (see Table \ref{A901table}).  }

\begin{figure}
\begin{center}
\includegraphics[clip,width=\columnwidth]{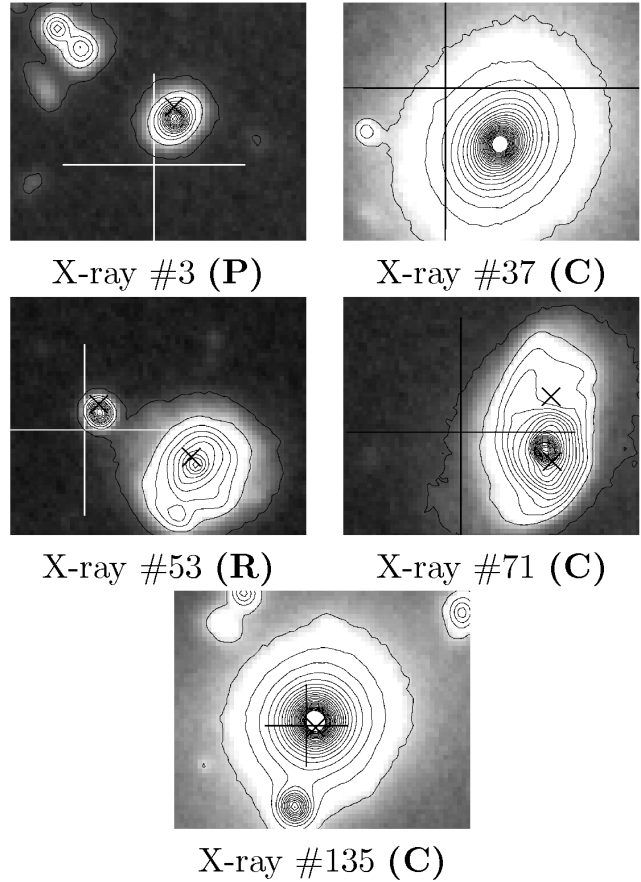}
\caption[Images of candidate supercluster AGN without spectra]{R-band
images of candidate supercluster AGN which were not observed with 2dF.
The scale and symbols are the same as in Figure 14.
}\label{NospecImages}
\end{center}
\end{figure}

\item{{\bf \# 20 -- supercluster AGN. } The 2dF spectrum of the
securely identified host galaxy shows weak emission lines. The X-ray
emission could be caused by a galaxy with a SFR of $\sim 100 {\rm
M}_{\odot}/{\rm yr}$ and no AGN but this is inconsistent with the SFR
implied by the [OII] line luminosity ($1.2 {\rm M}_{\odot}/{\rm yr}$)
and the nature of the galaxy in the 2dF spectrum. It is therefore
concluded that this galaxy contains an AGN which is optically
obscured. This conclusion is backed by the moderate X-ray luminosity
and X-ray to optical flux ratio of 0.12.}

\item{{\bf \# 24 -- supercluster AGN. } The host galaxy spectrum has
an upper limit on the SFR from [OII] of $< 1 {\rm M}_{\odot}/{\rm yr}$
which is 500 times less than that required to explain the X-ray
emission. Again, this must be due to optically obscured AGN
activity. The X-ray/Optical flux ratio also indicates an AGN. }

\item{{\bf \# 34 -- supercluster AGN. } The high hardness ratio makes
 AGN activity very likely, and the host galaxy is in the
 supercluster. The [OII] line flux gives a star-formation rate which
 is at least 25 times too low to account for the X-ray emission.  }



\begin{figure}
\begin{center}
\includegraphics[clip,width=\columnwidth,angle=0]{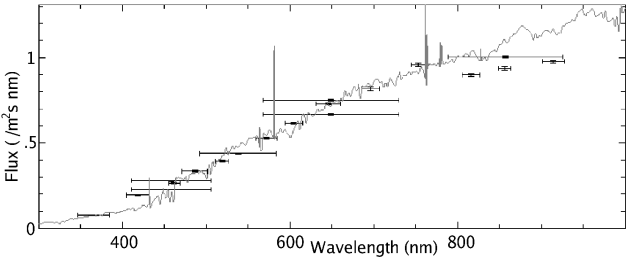}
\caption[COMBO-17 data and spectral template for object 12953]{The
COMBO-17 data and a z=0.16 galaxy template for optical object 12953,
associated with X-ray source \# 3. High levels of variability were
observed in the $B$ and $R$ bands, which were re-observed over
multiple runs, and which encompass the [OII] and [OIII] emission lines
at this redshift, indicating that the galaxy contains significant AGN
light and could be at $z \sim 0.16$. Above 800nm the data and template
do not match well as the illustrated template is not dust reddened -- a
bluer galaxy template with dust reddening would fit far
better. In addition the majority of galaxies with similar properties as
object 12953 are confirmed supercluster members. It is possible, but
by no means certain, that this is the correct redshift for this
object.
\label{specfit_12953}}
\end{center}
\end{figure}

\item{{\bf \# 37 -- supercluster AGN.} Moderate X-ray luminosity and
no detectable hard X-ray emission indicate a weak AGN or moderately
powerful starburst. The optical counterpart is an elliptical
supercluster galaxy but the probability of a match is only ~0.8 as the
error on the X-ray position is very large. This galaxy has a COMBO-17
`old red' fit to the SED, is the brightest galaxy in the south-west
group and falls on the main sequence of the colour magnitude
diagram. Other sources with similar properties but with optical
spectra ( in particular \#79 and \#24 ) are found to be
optically-obscured AGN. The galaxy was detected in the NVSS
\citep{NVSS} so is a radio source. The match with the radio data
secures the X-ray -- optical association, and makes it very probable
that this source is a supercluster AGN.}

\item{{\bf \# 53 -- rejected.} A soft, moderate luminosity ($3.8
\times 10^{41}$erg s$^{-1}$) X-ray source with two possible optical
counterparts, only one of which is in the supercluster. The
non-supercluster object is twice as likely to be the match by position
and luminosity only, and is also an optical quasar. Taking into
account the small number of bright quasars in this field, and the high
probability of them having detectable X-ray emission, the quasar is
highly unlikely to be a random association.}

\item{{\bf \# 71 -- supercluster AGN.} This source is securely
identified with a supercluster galaxy, without a 2dF spectrum. The high
hardness ratio and X-ray luminsosity indicate AGN activity.  }

\item{{\bf \# 79 -- supercluster AGN.}  The optical spectrum of the
supercluster host is that of a red early-type galaxy, whose
emission lines indicate at least 150 times too little star formation
to account for the X-ray emission.  }

\item{{\bf \# 81 -- supercluster AGN.} The X-ray emission is clearly
centred on a supercluster galaxy. The [OIII]/H$\beta$ ratio from the
2dF spectrum (taking an upper limit on the non detection of H$\beta$)
is 7.5, indicating that this source is likely to be an AGN. Despite
being prominent, the [OII] line flux is still 12 times too low to
account for the X-ray emission through star formation.  }

\item{{\bf \# 104 -- supercluster AGN. } This source is securely
identified with a supercluster galaxy. A hardness ratio of 0.95 and
$f_X/f_R$ of 0.22 make this an unambiguous AGN in the X-ray. }

\item{{\bf \# 105 -- supercluster AGN.}  This X-ray source is securely
identified with a supercluster galaxy. The  moderate X-ray luminosity
could not be caused by star-formation, as the optical spectrum is a
red galaxy with no detectable emission lines.  }

\item{{\bf \# 119 -- rejected.} The X-ray emission clearly emanates
from a supercluster galaxy. A hard X-ray spectrum, moderate
X-ray luminosity, but low $f_X/f_R$ appear to indicate an AGN or highly
obscured star-burst. The optical spectrum is clearly a blue
star-forming galaxy. However the observed [OII]
flux is not quite enough to account for all of the X-ray emission through
star-formation. To test whether this is due to a significant amount of
dust extinction the H$\beta$/H$\gamma$ ratio was calculated, using the
equivalent widths of the lines and the continuum flux density in the
$\lambda_{cen} = 5710$\AA \hspace{0.05cm} narrow-band filter to
calculate the flux from H$\beta$, and the mean flux densities in the
$\lambda_{cen} = 5190$\AA
\hspace{0.05cm}and $\lambda_{cen} = 4860$\AA \hspace{0.05cm} COMBO-17
bands to calculate the flux from H$\gamma$. The ratio is 1.35 times

\noindent lower than it should be, which can be accounted for by a E(B-V) dust
extinction of 0.677 magnitudes. Accounting for this, the [OII] SFR is
43 M$_\odot$/yr, which agrees (within errors) with the X-ray SFR of 30
M$_\odot$/yr. The source is clearly a star-forming supercluster
galaxy, so is rejected.}

\item{{\bf \# 127 -- rejected.} A hardness ratio of 0.25 indicates a
probable AGN. This source has two possible optical counterparts, with
the non-supercluster object being both ten times closer to the X-ray
position and a $m_{\rm{R}} = 22.4$ quasar. Accounting for the number
of luminous QSOs in the sample (90 with $m_{\rm R} < 22.4$), compared
to galaxies brighter than the possible supercluster host (285) the
quasar is the most probable counterpart by far.  However, the spectrum
of the supercluster object has [OII] and [OIII] emission lines, so it
is a possible source of X-rays from star-formation or an AGN. The
X-ray source appears to be a blend of two sources with an elongation
in the direction of the supercluster galaxy -- if this is the case
then the supercluster galaxy could contain a very faint AGN. However
the most likely source of the X-rays remains the QSO, and this source
is therefore rejected.}

\item{{\bf \# 135 -- supercluster AGN.} The X-ray spectrum of this
very luminous source is well fit by a power law, so must be an
AGN. The initial COMBO-17 photometric redshift of the host galaxy was
0.33, but the colours and position (a large, moderately red galaxy
near the centre of A901a) indicate supercluster membership. A visual
fit to a $z=0.16$ red early-type galaxy template is excellent above
4000\AA \hspace{0.05cm}as shown in Figure \ref{specfit_41435}. The
discrepancy can be accounted for by the variability of the source
(only this source and \#3 are variable by $>3\sigma$) and also by
adding a UV excess caused by the emission from the AGN. A spectrum of
the host galaxy was taken by \cite{BauerA901}, with the caveat that
the spectrum was ``poor quality'', and the redshift was given as
0.1585. The excellent optical match, spectroscopic redshift and
position of the emission relative to A901a lead to the conclusion that
this is a supercluster AGN. This source may be significantly
contaminated by optical AGN light.}


\addtocounter{figure}{+1}
\begin{figure}
\begin{center}
\includegraphics[clip,width=\columnwidth,angle=0]{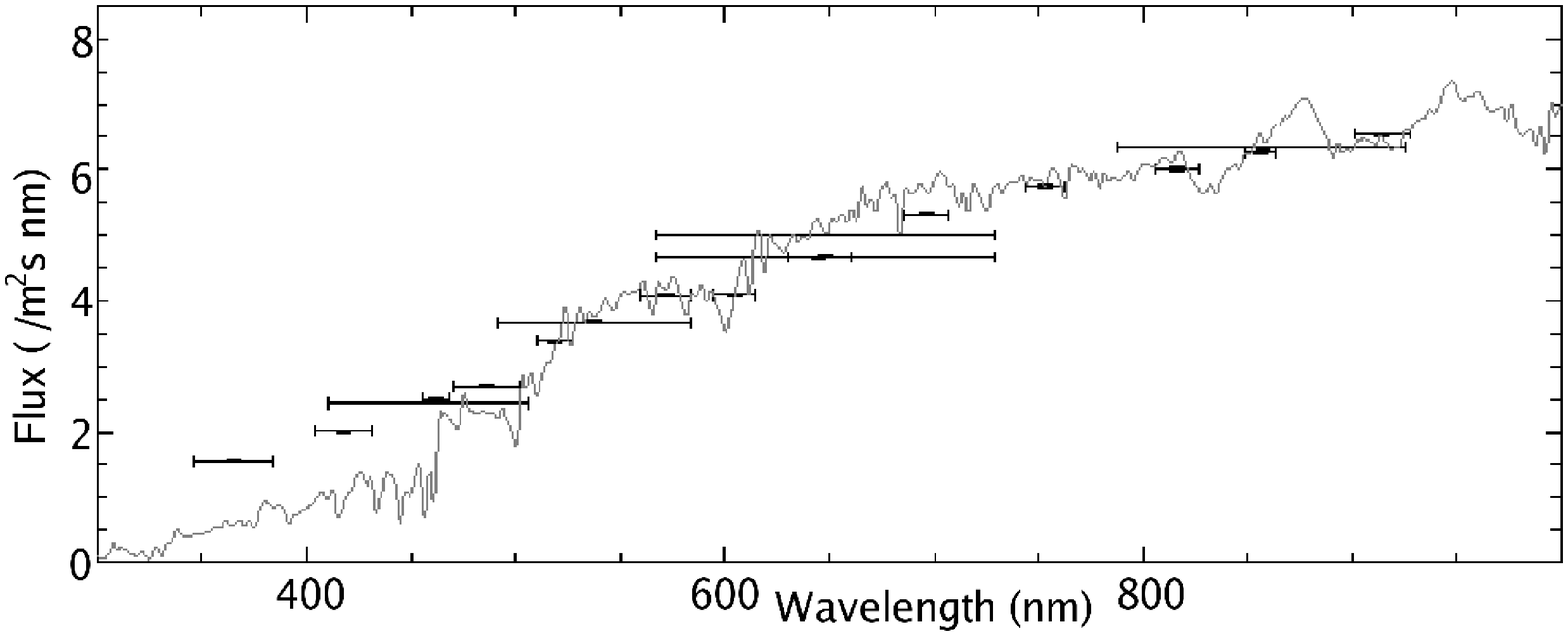}
\caption[COMBO-17 data and spectral template for object 41435]{The COMBO-17
data and a z=0.16 template for optical object 41435, associated with X-ray
source \# 135. The template has been selected by eye as a best guess
early-type template at this redshift. Adding excess UV light from an
obscured AGN would give a good fit, indicating that this source is likely
to be in the supercluster. The source is variable in the broad band at $>3\sigma$ significance.
\label{specfit_41435}}
\end{center}
\end{figure}

\item{{\bf \# 139 -- supercluster AGN. } A high hardness ratio and
X-ray luminosity, together with [OIII]$\lambda$5007/H$\beta > 20$ make
this a certain AGN, and it is securely associated with a supercluster
galaxy. }

\end{itemize}

\begin{figure}
\includegraphics[angle=270,scale=1.0]{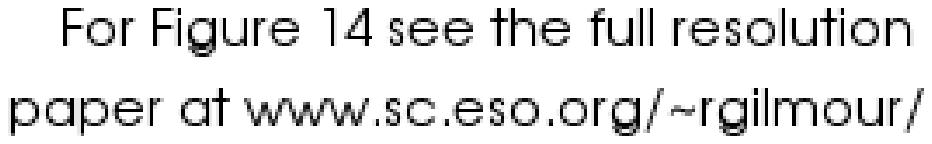}
\end{figure}

\begin{onecolumn}
\footnotesize
\setlength{\LTleft}{\fill}
\setlength{\tabcolsep}{0.12cm}
\begin{longtable}{rrrrrrrrrrrrrl}\\
\multicolumn{14}{c}{{\tablename} \thetable{}} \\[0.5ex]
\hline \hline \\[-2ex]
ID(X)&RA(X)$^{\ast}$ & Dec(X)$^{\ast}$ & Error$^{\star}$ & Rate$^{\sharp}$ & ID(Opt) & RA(Opt)$^{\ast}$ & Dec(Opt)$^{\ast}$ & $z_{\rm phot}$& Rel$^{\dag}$ & Prob$^{\ddag}$ & F(Sp)$^{\flat}$ & P(Sp)$^{\$}$& Notes\\ \hline
\endfirsthead

\multicolumn{14}{c}{{\tablename} \thetable{} -- Continued} \\[0.5ex]
\hline \hline \\[-2ex]
ID(X)&RA(X)$^{\ast}$ & Dec(X)$^{\ast}$ & Error$^{\star}$ & Rate$^{\sharp}$ & ID(Opt) & RA(Opt)$^{\ast}$ & Dec(Opt)$^{\ast}$ & $z_{\rm phot}$& Rel$^{\dag}$ & Prob$^{\ddag}$ & F(Sp)$^{\flat}$ & P(Sp)$^{\$}$& Notes\\ \hline
\endhead
\multicolumn{14}{r}{{Continued on next page\ldots}} \\
\endfoot
\\[-1.8ex] \hline \hline
\endlastfoot

1&09:56:36.7&-10:02:39.2&0.00185& 0.80& 28832  &9:56:36.5  &-10:02:43.5  & 0.253 & 0.852  &       0.852  & 0.39  & 0.79	& 	\\			 
2&09:56:35.9&-09:53:29.2&0.00095& 2.11& 48944  &9:56:35.7  &-09:53:28.7  & 0.769 & 0.850  &	  0.522  & 0.23  & 0.65	& A 	\\		 
2&&&& &				        48703  &9:56:35.8  &-09:53:30.0  & 0.759 & 0.808  &	  0.386  & 0.23  & 0.22	& A 	\\		 
3&09:56:35.8&-10:10:12.8&0.00132& 5.17& 12953  &9:56:35.7  &-10:10:10.6  &  -    & 0.940  &	  0.842  & 5.16  & 0.92	& Z(a)	\\		 
4&09:56:31.2&-09:53:28.9&0.00102& 2.44&&&&&                                                              & 0.22  & 0.95 &       \\   	  	     		   	 
5&09:56:28.0&-10:01:08.3&0.00155& 0.75& 32205  &9:56:28.6  &-10:01:07.8  & x     & 0.764  &	  0.605  &       &   	& Star	\\			 
5&&&& &				        32229  &9:56:28.0  &-10:01:03.5  & 1.301 & 0.526  &	  0.207  &       &   	& QSO	\\		 
6&09:56:20.4&-10:05:22.8&0.00113& 1.21& 23122  &9:56:20.3  &-10:05:24.4  & 0.616 & 0.880  &	  0.736  &       &   	& 	\\			 
7&09:56:15.3&-09:53:19.0&0.00135& 0.58&&&&&&&&\\    					   	  	   	     	   	 
8&09:56:12.1&-09:50:44.0&0.00112& 5.06& 54851  &9:56:11.9  &-09:50:47.2  & 1.353 & 0.904  &	  0.904  & 1.00  & 0.91	& QSO	\\			 
9&09:56:08.7&-10:08:03.0&0.00093& 0.82&&&&&&&&\\    					   	  	   	     	   	 
10&09:57:01.8&-09:55:27.5&0.00241& 0.85&&&&&&&&\\   					   	  	   	     	   	 
11&09:56:41.0&-09:52:49.8&0.00123& 0.77& 51604  &9:56:40.4  &-09:52:49.8  & 0.164 & 0.849  &	  0.681  & 0.12  & 0.08 & B Spec\\		 
11&&&& &			         50205  &9:56:41.2  &-09:52:51.9  & 1.898 & 0.622  &	  0.198  & 0.19  & 0.61 & B QSO	\\		 
12&09:56:10.7&-10:16:05.7&0.00129& 1.04&&&&&&&&\\   					   	  	   	     	   	 
13&09:56:01.4&-10:00:25.6&0.00070& 1.90&&&&&&&&\\   					   	  	   	     	   	 
14&09:55:32.3&-09:58:54.8&0.00164& 0.56& 39465  &9:55:31.9  &-09:58:59.6  & x     & 0.999  &	  0.999  &       &   	& Star 	\\			 
15&09:55:31.6&-10:06:11.4&0.00289& 0.73&&&&&&&&\\   					   	  	   	     	   	 
16&09:55:28.2&-09:58:59.6&0.00217& 0.63& 36617  &9:55:28.1  &-09:59:15.5  & x     & 0.815  &	  0.563  &       &   	& Star	\\		 
16&&&& &				 36776  &9:55:27.7  &-09:58:58.7  & x     & 0.625  &	  0.212  &       &   	& Star	\\		 
17&09:56:54.1&-10:02:49.0&0.00085& 2.85& 28744  &9:56:54.0  &-10:02:45.9  & 0.754 & 0.941  &	  0.923  & 1.19  & 0.91	& 	\\			 
18&09:56:47.3&-10:13:29.3&0.00087& 1.53& 5514   &9:56:47.1  &-10:13:30.1  & 1.201 & 0.972  &	  0.972  & 1.23  & 0.93	& QSO	\\			 
19&09:56:20.0&-10:01:17.0&0.00116& 0.40&&&&&&&&\\   					   	  	   	     	   	 
20&09:56:17.7&-10:01:49.3&0.00107& 1.55& 31178  &9:56:17.6  &-10:01:49.4  & 0.171 & 0.982  &	  0.982  & 1.93  & 0.90	& Spec	\\		 
21&09:56:10.3&-09:58:59.4&0.00080& 2.27& 36827  &9:56:10.2  &-09:58:59.2  & 1.886 & 0.953  &	  0.906  & 0.54  & 0.90	& QSO	\\			 
22&09:55:57.6&-10:01:27.5&0.00060& 2.18& 31519  &9:55:57.5  &-10:01:28.3  & 2.413 & 0.992  &	  0.987  & 1.18  & 0.93	& QSO	\\			 
23&09:55:52.6&-09:59:51.1&0.00060& 6.77& 35608  &9:55:52.5  &-09:59:51.3  & x     & 0.998  &	  0.998  &       & 	& Star	\\			 
24&09:55:44.2&-09:59:33.0&0.00100& 1.88& 36966  &9:55:44.3  &-09:59:33.5  & 0.175 & 0.982  &	  0.934  &       &   	& Spec	\\		 
25&09:55:41.7&-09:59:20.9&0.00127& 0.77&&&&&&&&\\  					   	  	   	     	   	 
26&09:55:39.4&-10:13:25.9&0.00168& 1.18&&&&&&&&\\		 
27&09:55:34.6&-09:56:01.6&0.00094& 4.24& 43454  &9:55:34.6  &-09:56:4.17  & 1.655 & 0.977  &	  0.962  & 0.94  & 0.90	& QSO	\\			 
28&09:56:33.6&-09:53:55.1&0.00134& 1.29& 47810  &9:56:33.6  &-09:53:58.4  & 2.109 & 0.881  &	  0.881  & 0.42  & 0.83	& QSO	\\			 
29&09:56:01.3&-10:06:40.0&0.00138& 0.57&&&&&&&&\\  					   	  	   	     	   	 
30&09:56:00.1&-10:09:03.5&0.00121& 0.57&&&&&&&&\\  					   	  	   	     	   	 
31&09:56:00.1&-09:55:32.8&0.00142& 0.73&&&&&&&&\\  					   	  	   	     	   	 
32&09:55:50.0&-09:59:44.8&0.00110& 1.85&&&&&&&&\\  					   	  	   	     	   	 
33&09:56:55.4&-10:02:18.0&0.00142& 0.69&&&&&&&&\\  					   	  	   	     	   	 
34&09:56:48.2&-09:58:03.0&0.00156& 0.85& 39549  &9:56:48.1  &-09:58:1.65  & 0.166 & 0.962  &	  0.962  &       & 	& Spec	\\		 
35&09:56:05.3&-09:51:52.6&0.00154& 1.01&&&&&&&&\\  					   	  	   	     	   	 
36&09:55:53.6&-10:14:11.0&0.00296& 0.84&&&&&&&&\\  					   	  	   	     	   	 
37&09:55:38.6&-10:10:15.9&0.00416& 0.99& 14161  &9:55:38.4  &-10:10:19.1  & 0.171 & 0.809  &	  0.759  &       &   	& 	\\			 
38&09:56:27.5&-10:08:19.5&0.00767& 0.60&&&&&&&&\\   					   	  	   	     	   	 
39&09:57:07.4&-09:56:48.4&0.00060& 1.77& 42260  &9:57:07.2  &-09:56:44.9  & x     & 0.975  &	  0.975  &       &   	& Star	\\			 
40&09:57:03.6&-09:55:04.7&0.00123& 1.88&        &           &             &       &        &	         &       & 	&  	\\			 
41&09:57:00.7&-09:58:29.6&0.00146& 0.90&&&&&&&&\\   					   	  	   	     	   	 
42&09:57:00.6&-09:54:24.1&0.00117& 2.17&&&&&&&&\\    					   	  	   	     	   	 
43&09:56:58.4&-10:10:29.8&0.00120& 1.22& 12232  &9:56:58.4  &-10:10:28.1  & 3.448 & 0.924  &	  0.924  & 0.46  & 0.91	& QSO	\\			 
44&09:56:56.3&-09:54:19.8&0.00123& 1.45&&&&&&&&\\   					   	  	   	     	   	 
45&09:56:55.6&-09:55:07.5&0.00096& 1.48&&&&&&&&\\   					   	  	   	     	   	 
46&09:56:49.5&-10:07:24.5&0.00195& 1.21&&&&&&&&\\   					   	  	   	     	   	 
47&09:56:47.4&-10:02:34.7&0.00123& 1.07&&&&&&&&\\			 
48&09:56:43.8&-09:55:40.0&0.00105& 1.09& 44635  &9:56:43.9  &-09:55:39.9  & 0.083 & 0.984  &	  0.597  & 20.80 & 0.26	& C	\\		 
48&&&& &				 45154  &9:56:43.9  &-09:55:43.2  & 0.053 & 0.976  &	  0.391  & 23.36 & 0.45	& C	\\		 
49&09:56:42.4&-10:13:11.6&0.00100& 1.47& 6258   &9:56:42.3  &-10:13:11.2  & 0.337 & 0.951  &	  0.913  & 0.85  & 0.91	& 	\\			 
50&09:56:41.9&-10:08:50.5&0.00108& 0.90& 15780  &9:56:42.0  &-10:08:48.9  & 1.327 & 0.851  &	  0.761  & 0.14  & 0.78 & QSO	\\			 
51&09:56:42.2&-10:05:58.6&0.00138& 0.56& 21892  &9:56:42.1  &-10:05:56.0  & 2.267 & 0.765  &	  0.751  & 0.38  & 0.83	& 	\\			 
52&09:56:40.8&-09:59:16.4&0.00092& 0.44& 36062  &9:56:40.9  &-09:59:16.0  & 0.728 & 0.762  &	  0.541  &       &   	&	\\			 
52&&&& &				 36147  &9:56:40.7  &-09:59:14.2  & x     & 0.631  &	  0.289  &       &   	&	\\			 
53&09:56:40.6&-10:11:49.7&0.00124& 1.54& 9081   &9:56:40.6  &-10:11:49.1  & 1.290 & 0.938  &	  0.639  & 0.68  & 0.65	& QSO	\\		 
53&&&& &				 9524   &9:56:40.2  &-10:11:52.0  & 0.176 & 0.883  &	  0.318  & 0.57  & 0.23	& 	\\		 
54&09:56:40.6&-10:00:30.4&0.00089& 0.75&&&&&&&&\\    					   	  	   	     	   	 
55&09:56:40.0&-10:09:30.1&0.00100& 1.45& 14419  &9:56:39.8  &-10:09:30.1  & 0.292 & 0.939  &	  0.915  &       &   	& 	\\			 
56&09:56:39.6&-10:09:00.3&0.00126& 0.72&&&&&&&&\\   					   	  	   	     	   	 
57&09:56:39.7&-09:57:18.3&0.00086& 1.12& 46236  &9:56:39.6  &-09:57:17.5  & x     & 1.000  &	  1.000  & 0.59  & 0.99	& 	\\			 
58&09:56:37.3&-10:03:16.2&0.00070& 2.05& 27507  &9:56:37.3  &-10:03:17.1  & 1.458 & 0.981  &	  0.973  & 0.48  & 0.91	& QSO	\\			 
59&09:56:37.0&-09:52:37.2&0.00079& 3.36& 50887  &9:56:37.0  &-09:52:37.6  & 0.376 & 0.975  &	  0.937  & 0.55  & 0.91	& 	\\			 
60&09:56:36.1&-10:01:49.9&0.00074& 1.31& 30570  &9:56:36.1  &-10:01:51.3  & 0.948 & 0.925  &	  0.753  & 0.76  & 0.88	& 	\\			 
61&09:56:35.6&-10:00:04.3&0.00097& 0.34& 34255  &9:56:35.9  &-10:00:05.9  & 1.667 & 0.763  &	  0.386  &       &   	& QSO	\\		 
61&&&& &				  34746  &9:56:36.0  &-09:59:57.8  & x     & 0.631  &	  0.204  & 0.11  & 0.74 & Star	\\		 
61&&&& &				  34238  &9:56:35.3  &-10:00:09.0  & 0.426 & 0.631  &	  0.204  &       &   	&	\\			 
62&09:56:35.3&-10:04:55.2&0.00060& 55.10& 24409  &9:56:35.3  &-10:04:54.6  & x     & 0.993  &	  0.993  &       & 	& QSO	\\			 
63&09:56:34.5&-09:59:30.1&0.00098& 1.10& 35643  &9:56:34.5  &-09:59:30.1  & 0.965 & 0.952  &	  0.952  &       &	& QSO	\\			 
64&09:56:30.6&-10:00:16.4&0.00060& 38.90& 36653  &9:56:30.6  &-10:00:15.5  & 0.000 & 1.000  &     1.000  & 0.66  & 1.00& Star	\\			 
65&09:56:29.9&-09:52:37.4&0.00167& 0.31&&&&&&&&\\  					   	  	   	     	   	 
66&09:56:29.7&-10:02:01.3&0.00067& 1.87& 63777$^1$  &9:56:29.6  &-10:01:59.7  & x     & 0.952   & 0.952  & 0.82  & 0.99    & 	\\			 
67&09:56:29.1&-10:10:05.2&0.00093& 1.13&&&&&&&&\\   					   	  	   	     	   	 
68&09:56:29.1&-09:51:33.0&0.00130& 0.94&&&&&&&&\\			 
69&09:56:26.7&-10:05:10.0&0.00060& 3.74& 23665  &9:56:26.8  &-10:05:09.5  & 0.601 & 0.990  &	  0.497  &       & 	& QSO	\\			 
69&09:56:26.7&-10:05:10.0&0.00060& 3.74& 24046  &9:56:27.1  &-10:05:10.2  & x     & 0.990  &	  0.497  &       & 	& Star	\\			 
70&09:56:26.5&-10:03:24.6&0.00119& 0.69& 27308  &9:56:26.5  &-10:03:22.1  & 0.924 & 0.814  &	  0.500  &       & 	& 	\\			 
70&09:56:26.5&-10:03:24.6&0.00119& 0.69& 27176  &9:56:27.0  &-10:03:26.4  & 0.179 & 0.627  &	  0.192  &       & 	& 	\\			 
70&09:56:26.5&-10:03:24.6&0.00119& 0.69& 27071  &9:56:26.7  &-10:03:26.2  & 0.903 & 0.628  &	  0.192  &       & 	& 	\\			 
71&09:56:26.4&-10:10:57.9&0.00166& 0.68& 11827  &9:56:26.1  &-10:10:58.5  & 0.175 & 0.938  &	  0.938  & 2.88  & 0.76	& D 	\\		 
72&09:56:26.5&-09:55:55.3&0.00110& 1.34& 46335  &9:56:26.5  &-09:55:46.4  & x     & 0.998  &	  0.980  &       & 	& Star	\\			 
73&09:56:22.6&-09:56:00.6&0.00066& 1.60& 43383  &9:56:22.5  &-09:56:00.0  & 1.253 & 0.814  &	  0.814  & 0.11  & 0.72	& 	\\			 
74&09:56:21.7&-10:03:06.8&0.00065& 0.96& 27810  &9:56:21.7  &-10:03:06.1  & 2.000 & 0.952  &	  0.952  & 0.20  & 0.84 & QSO	\\			 
75&09:56:21.2&-09:56:36.1&0.00093& 1.14& 42064  &9:56:21.3  &-09:56:37.7  & 3.493 & 0.927  &	  0.927  & 0.09  & 0.82	& QSO	\\			
76&09:56:20.1&-10:03:50.3&0.00061& 1.84& 26320  &9:56:20.1  &-10:03:48.6  & 0.986 & 0.883  &	  0.883  & 0.09  & 0.74 & 	\\			 
77&09:56:20.0&-10:00:48.5&0.00090& 0.54& 32961	&9:56:19.9  &-10:00:46.4  & 0.908 & 0.760  &      0.652  & 0.22  & 0.81 & E	\\
78&09:56:19.7&-10:03:27.1&0.00063& 0.85& 27050  &9:56:19.8  &-10:03:27.1  & 0.974 & 0.814  &	  0.814  & 0.09  & 0.66	& 	\\			 
79&09:56:18.8&-09:55:57.9&0.00139& 1.06& 44351  &9:56:18.8  &-09:55:57.8  & 0.162 & 0.980  &	  0.980  &       &   	& Spec	\\		 
80&09:56:18.1&-09:53:59.8&0.00078& 16.40& 47978  &9:56:18.0  &-09:54:01.2  & 1.133 & 0.992  &	  0.992  & 4.65  & 0.94	& QSO	\\			 
81&09:56:17.7&-10:07:20.2&0.00126& 0.42& 19305  &9:56:17.7  &-10:07:18.7  & 0.175 & 0.970  &	  0.960  & 0.46  & 0.92	& Spec	\\		 
82&09:56:15.9&-10:02:19.2&0.00060& 0.21&&&&&&&&\\   					   	  	   		   	 
83&09:56:15.5&-09:50:31.3&0.00150& 0.66&&&&&&&&\\   					   	  	   		   	 
84&09:56:15.0&-09:58:20.6&0.00060& 4.38& 38125  &9:56:15.0  &-09:58:21.3  & 0.902 & 0.851  &	  0.851  & 0.84  & 0.72	& 	\\			 
85&09:56:13.1&-09:59:03.8&0.00114& 0.44& 36939  &9:56:13.1  &-09:59:04.9  & x     & 0.987  &	  0.987  &       &   	& Star 	\\			 
86&09:56:13.0&-10:04:07.3&0.00060& 2.00& 25718  &9:56:13.0  &-10:04:06.9  & 2.267 & 0.969  &	  0.969  & 0.10  & 0.87	& QSO	\\			 
87&09:56:11.8&-09:59:55.9&0.00100& 0.65&&&&&&	                                                           0.12  & 0.92 &       \\  	   	     	   	 
88&09:56:10.6&-09:49:12.0&0.00101& 2.44& 58384  &9:56:10.5  &-09:49:13.2  & 1.899 & 0.951  &	  0.835  & 1.31  & 0.92	& QSO	\\			 
89&09:56:10.0&-10:07:11.3&0.00086& 0.72& 19716  &9:56:10.1  &-10:07:10.8  & 0.260 & 0.975  &	  0.975  & 0.25  & 0.90	& 	\\			 
90&09:56:05.5&-10:00:30.0&0.00068& 1.11& 33381  &9:56:05.4  &-10:00:30.5  & 0.961 & 0.764  &	  0.508  & 1.11  & 0.71	& F 	\\	
91&09:56:03.3&-10:07:41.1&0.00087& 2.25& 18874  &9:56:03.3  &-10:07:42.4  & 0.255 & 0.987  &	  0.983  & 6.53  & 0.93	& 	\\			 
92&09:56:00.9&-09:56:18.7&0.00251& 2.31&&&&&&&&\\    					   	  	   	     	   	 
93&09:55:58.2&-10:07:27.4&0.00101& 1.40& 18722  &9:55:58.3  &-10:07:27.3  & 1.304 & 0.939  &	  0.882  & 0.67  & 0.80	& QSO	\\			 
94&09:55:57.9&-10:06:52.7&0.00121& 1.40& 19933  &9:55:57.9  &-10:06:54.1  & 0.838 & 0.950  &	  0.863  & 0.46  & 0.91	& 	\\			 
95&09:55:55.7&-10:08:30.7&0.00115& 0.65& 16610  &9:55:55.7  &-10:08:31.9  & 0.516 & 0.906  &	  0.782  & 1.46  & 0.90	& 	\\			 
96&09:55:54.6&-10:06:46.5&0.00134& 0.49&&&&&&&&\\    					   	  	   	     	   	 
97&09:55:53.7&-10:08:47.9&0.00162& 0.39&&&&&&&&\\    					   	  	   	     	   	 
98&09:55:52.9&-10:05:43.7&0.00110& 0.82& 22404  &9:55:52.9  &-10:05:43.9  & 0.445 & 0.938  &	  0.938  & 0.43  & 0.91	&  	\\			 
99&09:55:52.6&-10:06:44.6&0.00121& 0.43&&&&&&&&\\    					   	  	   	     	   	 
100&09:55:52.5&-10:04:28.7&0.00074& 1.88& 24884  &9:55:52.5  &-10:04:28.8  & 0.416 & 0.942  &	  0.942  & 0.38  & 0.88	& 	\\			 
101&09:55:51.2&-10:03:56.7&0.00133& 1.04&&&&&&&&\\   					   	  	   	     	   	 
102&09:55:50.5&-09:52:06.2&0.00137& 2.12& 52252  &9:55:50.4  &-09:52:07.3  & 0.240 & 0.923  &	  0.693  &       &   	&	\\			 
102&&&& &				  53126  &9:55:50.6  &-09:51:53.1  & x     & 0.813  &	  0.249  &       &   	& Star	\\		 
103&09:55:43.9&-09:57:05.3&0.00148& 0.74&&&&&                                                            & 0.26  & 0.94 &       \\   	 
104&09:55:43.6&-10:09:09.5&0.00113& 2.03& 15698  &9:55:43.7  &-10:09:11.8  & 0.179 & 0.961  &	  0.387  & 0.73  & 0.92	& G Spec\\	
105&09:55:43.6&-10:08:28.1&0.00117& 0.62& 17675  &9:55:43.4  &-10:08:26.5  & 0.170 & 0.970  &	  0.844  & 0.30  & 0.81	& Spec	\\		 
106&09:55:38.1&-10:08:25.3&0.00112& 1.44& 16625  &9:55:38.2  &-10:08:24.6  & 0.562 & 0.812  &	  0.812  & 0.15  & 0.79	& 	\\			 
107&09:55:36.9&-09:57:15.8&0.00114& 2.17& 40605  &9:55:36.8  &-09:57:17.1  & 0.975 & 0.924  &	  0.809  & 0.65  & 0.91	& QSO	\\			 
108&09:55:35.8&-10:09:15.8&0.00075& 2.56& 14758  &9:55:35.9  &-10:09:15.7  & 1.693 & 0.942  &	  0.938  & 0.13  & 0.85	& QSO	\\			 
109&09:55:35.1&-10:01:52.7&0.00102& 0.59& 30592  &9:55:35.0  &-10:01:51.5  & 0.814 & 0.811  &	  0.286  & 0.33  & 0.74	& H	\\			 
109&          &           &       &     & 31111  &9:55:34.7  &-10:01:52.7  & 0.249 & 0.882  &	  0.493  &       & 	& H	\\			 
110&09:55:32.3&-10:01:44.9&0.00126& 0.96& 30855  &9:55:32.6  &-10:01:49.0  & x     & 0.961  &	  0.845  &       &   	& Star	\\			 
111&09:55:31.1&-10:05:22.5&0.00163& 0.60& 28232  &9:55:30.9  &-10:05:18.9  & x     & 1.000  &	  1.000  & 0.41  & 0.87	& 	\\			 
112&09:55:28.5&-10:05:30.8&0.00060& 7.21& 22807  &9:55:28.6  &-10:05:32.2  & 0.257 & 0.981  &	  0.981  & 0.91  & 0.92	& 	\\			 
113&09:57:18.5&-09:59:51.5&0.00109& 0.98&&&&&&&&\\   					   	  	   	     	   	 
114&09:57:11.3&-09:57:05.5&0.00140& 1.17&&&&&&&&\\  					   	  	   	     	   	 
115&09:56:57.1&-10:05:43.7&0.00198& 0.44&&&&&&&&\\  					   	  	   	     	   	 
116&09:56:52.4&-10:00:30.4&0.00124& 1.11& 35364  &9:56:52.2  &-10:00:29.7  & 0.079 & 0.984  &	  0.984  & 5.45  & 0.99	& D	\\		 
117&09:56:48.1&-09:53:36.4&0.00139& 0.76&&&&&&&&\\   					   	  	   	     	   	 
118&09:56:47.1&-10:11:04.2&0.00110& 1.02& 10744  &9:56:47.2  &-10:11:03.8  & 0.941 & low 	  &low   & 0.34  & 0.42	& I	\\			 
119&09:56:46.7&-10:01:37.5&0.00156& 0.58& 31772  &9:56:46.7  &-10:01:38.6  & 0.168 & 0.969  &	  0.969  & 7.36  & 0.93	& Spec	\\		 
120&09:56:44.2&-09:50:46.6&0.00253& 0.99&&&&&&&&\\   					   	  	   	     	   	 
121&09:56:42.3&-10:12:18.4&0.00211& 0.59& 8770   &9:56:42.2  &-10:12:19.1  & 0.324 & 0.883  &	  0.482  &       &   	&	\\			 
121&&&& &				  7961   &9:56:42.6  &-10:12:22.6  & x     & 0.852  &	  0.369  &       &   	& Star	\\		 
122&09:56:24.8&-09:51:17.1&0.00157& 0.62&&&&&&&&\\   	   					   	  	     		   	 
123&09:56:23.5&-09:51:54.8&0.00117& 0.71&&&&&&&&\\   	   					   	  	     		   	 
124&09:56:22.0&-10:01:13.7&0.00109& 0.33& 31850  &9:56:22.0  &-10:01:14.5  & 3.373 & 0.882  &	  0.882  & 0.30  & 0.87	& QSO	\\			 
125&09:56:08.1&-10:06:12.6&0.00129& 0.30&&&&&&&&\\   					   	  	   	     	   	 
126&09:56:06.2&-10:10:30.7&0.00170& 0.30&&&&&&&&\\   					   	  	   	     	   	 
127&09:56:00.9&-10:08:24.2&0.00131& 0.33& 16637  &9:56:00.9  &-10:08:23.3  & 1.967 & 0.706  &	  0.406  & 0.10  & 0.67 & QSO	\\		 
127&&&& &				  17155  &9:56:1.51  &-10:08:19.1  & 0.173 & 0.631  &	  0.289  &       &   	& Spec	\\ 		 
128&09:55:58.8&-10:05:57.3&0.00124& 0.38&&&&&&&&\\   	   					   	  	     		   	 
129&09:55:42.8&-10:04:37.7&0.00128& 0.43&&&&&&&&\\   	   					   	  	     		   	 
130&09:55:39.1&-09:54:33.8&0.00155& 0.47&&&&&                                                            & 0.15  & 0.88 &       \\ 
131&09:55:31.2&-10:04:07.4&0.00137& 0.78&&&&&&&&\\   	   					   	  	     		   	 
132&09:57:10.6&-09:56:27.3&0.00118& 0.66& 42725  &9:57:10.4  &-09:56:24.0  & x     & 0.938  &	  0.938  &  1.03 & 0.91	& QSO	\\			 
133&09:56:12.0&-10:06:13.8&0.00138& 0.27&&&&&&&&\\   	   					   	  	     		   	 
134&09:56:10.4&-10:09:44.7&0.00214& 0.33&&&&&&&&\\   					   	  	   	     	   	 
135&09:56:28.2&-09:57:19.3&0.00060& 277.00& 41435  &9:56:28.2  &-09:57:19.0  & -     & 0.995  &	  0.966  & 1.15  & 0.94	& Z(b)	\\		 
136&09:56:21.8&-10:04:34.9&0.00064& 2.53& 24683  &9:56:21.8  &-10:04:35.7  & 1.088 & 0.968  &	  0.968  &       & 	& QSO	\\			 
137&09:56:18.8&-10:07:45.6&0.00101& 1.95& 18034  &9:56:18.7  &-10:07:47.0  & 0.834 & 0.850  &	  0.840  & 1.69  & 0.86	& 	\\			 
138&09:56:21.8&-10:06:48.2&0.00091& 0.42& 20029  &9:56:21.8  &-10:06:47.4  & 0.719 & 0.880  &	  0.880  & 0.15  & 0.82	& QSO	\\			 
139&09:56:50.2&-10:11:51.8&0.00215& 1.15& 9020   &9:56:50.0  &-10:11:55.8  & 0.171 & 0.851  &	  0.851  & 3.58  & 0.92	& Spec	\\ 
\hline        \\

\multicolumn{14}{p{\columnwidth}}{\bf Notes:}\\
\multicolumn{14}{p{\columnwidth}}{$\ast$ All co-ordinates are in J2000.}\\
\multicolumn{14}{p{\columnwidth}}{$\star$ Error on X-ray position in degrees, as described in Section \ref{point_src_prop}.}\\
\multicolumn{14}{p{\columnwidth}}{$\sharp$ Counts per second in units of $10^{-3}$ in the 0.5 - 7.5 keV band, given by the average of the {\sc wavdetect} count rates from the three images. The count rates from the PN detector were scaled down by a factor of 3 (calculated from the brightest sources) to match those from the MOS detectors. }\\
\multicolumn{14}{p{\columnwidth}}{$\dag$ Reliability; the probability that the X-ray -- optical association is not random.}\\
\multicolumn{14}{p{\columnwidth}}{$\ddag$ Probability; the chance that this X-ray -- optical association gives the true counterpart, given all possible associations and the chance of no counterpart existing. In the case of sources with more than one reliable match this is much lower than the individual reliabilities.}\\
\multicolumn{14}{p{\columnwidth}}{$\flat$ 24$\mu$m flux in mJy, from Spitzer data.}\\
\multicolumn{14}{p{\columnwidth}}{$\$$ Probability from 24$\mu$m data, given by the combined probability of the X-ray - 24$\mu$m match being unique, and the 24$\mu$m -- optical match. If no optical match is found then this is purely the X-ray -- 24$\mu$m probability.}\\
\multicolumn{14}{p{\columnwidth}}{x -- The source is too faint to have a reasonable COMBO redshift, or is a star.}\\
\multicolumn{14}{p{\columnwidth}}{1 -- This source was found by eye and added to the COMBO catalogue, with a R-band magnitude of 23.7.}\\
\multicolumn{14}{p{\columnwidth}}{A -- The X-ray source matches well with one 24$\mu$m
source. Both the X-ray to optical and 24$\mu$m to optical matches indicate that the first optical source is most likely to be the true counterpart.}\\
\multicolumn{14}{p{\columnwidth}}{B -- The 24$\mu$m matches are low probability, but are included here as one of the possible counterparts is a supercluster galaxy.}\\
\multicolumn{14}{p{\columnwidth}}{C --The two optical and two 24$\mu$m detections are both components of a local merging system. }\\
\multicolumn{14}{p{\columnwidth}}{D -- The  24$\mu$m catalogue includes multiple detections of the same object. The combined flux is given.}\\
\multicolumn{14}{p{\columnwidth}}{E -- The X-ray -- optical match has low probability, but it becomes more likely once the 24$\mu$m data is added.}\\
\multicolumn{14}{p{\columnwidth}}{F -- The 24$\mu$m data reduces 2 optical possibilities to one. The 24$\mu$m--X-ray match is excellent (0.99) but the match with the optical source is lower significance as the source is very faint.}\\ 
\multicolumn{14}{p{\columnwidth}}{G -- 24$\mu$m data reduces two equally likely X-ray -- optical matches to one secure match. The rejected match is a very bright star, which is actually $2\sigma$ away.}\\
\multicolumn{14}{p{\columnwidth}}{H -- The 24$\mu$m source is well matched to the optical counterpart, and is likely to be the source of the X-rays.}\\
\multicolumn{14}{p{\columnwidth}}{I -- The X-ray -- optical match gives a number of possible counterparts. All are rejected, as the 24$\mu$m source is an excellent match with the X-ray (0.98) and is not associated with any of these sources. The most likely match is a very faint source, which is listed.}\\
\multicolumn{14}{p{\columnwidth}}{QSO -- The optical source is a QSO, and therefore
has a higher likelihood of being the true counterpart than indicated by the probabilities.}\\
\multicolumn{14}{p{\columnwidth}}{Star -- The optical source is a star, and as stars
are often bright (and rare, giving a high probability) but not often X-ray
loud the source may be a chance association.}\\
\multicolumn{14}{p{\columnwidth}}{Spec -- This source has a 2dF spectrum}\\
\multicolumn{14}{p{\columnwidth}}{Z -- Dubious redshift, see Section \ref{A901_AGN_det}; Z(a) -- COMBO-17 redshift is 1.4 but it may be 0.16. Z(b) -- COMBO-17 redshift is 0.33, but $\sim$0.16 is found spectroscopically.}\\

\hline
\caption{X-ray sources and possible optical matches and 24$\mu$m fluxes.}\label{A901table}
\end{longtable}
\normalsize


\end{onecolumn}

\label{lastpage}

\end{document}